\documentclass[sigconf, screen, nonacm]{acmart}

\usepackage{algorithmic}
\usepackage{graphicx}
\usepackage{textcomp}
\usepackage{xcolor}
\usepackage{enumitem}
\usepackage{tabularx}
\usepackage{tikz}

\usepackage[final]{coauthorcomments}
\newcoauthor{sj}{SJ}{magenta}
\newcoauthor{hf}{HF}{orange}
\newcoauthor{my}{MY}{teal}
\newcoauthor{jt}{JT}{pink}
\newcoauthor{cw}{CW}{red}
\newcoauthor{fl}{FL}{purple}
\newcoauthor{xa}{CA}{brown}

\newcommand*\circled[1]{\smash{\tikz[baseline=(char.base)]{
    \node[inner sep=0pt, font=\footnotesize,
          text height=1.6ex, text depth=0.25ex] (char) {#1};
    \draw (char.center) circle (0.6em);}}}
\newcounter{observation}
\newenvironment{observation}
  {\refstepcounter{observation}\par\smallskip\noindent
   \begin{tabular}{|p{\dimexpr\columnwidth-2\tabcolsep-2\arrayrulewidth\relax}|}\hline
   \small\textbf{Observation~\theobservation.}\ }
  {\\\hline\end{tabular}\par\smallskip}

\newcommand{\FIXME}[1]{}

\begin{document}

\title{TripleKV: Predict, Prefetch, Protect -- KV Cache Management for AI Agents}

%% ---- Authors ----
% Four author blocks per row under the title.
\settopmatter{authorsperrow=4}

\author{Shashwat Jaiswal}
\affiliation{%
  \institution{University of Illinois Urbana-Champaign}
  \city{Urbana}
  \state{Illinois}
  \country{USA}}
\email{sj74@illinois.edu}

\author{Chloe Alverti}
\affiliation{%
  \institution{National Technical University of Athens}
  \city{Athens}
  \country{Greece}}
\email{xalverti@cslab.ece.ntua.gr}

\author{Mengmei Ye}
\affiliation{%
  \institution{IBM Research}
  \city{Yorktown Heights}
  \state{New York}
  \country{USA}}
\email{mye@ibm.com}

\author{Chen Wang}
\affiliation{%
  \institution{IBM Research}
  \city{Yorktown Heights}
  \state{New York}
  \country{USA}}
\email{chen.wang1@ibm.com}

\author{Hubertus Franke}
\affiliation{%
  \institution{IBM Research}
  \city{Yorktown Heights}
  \state{New York}
  \country{USA}}
\email{frankeh@us.ibm.com}

\author{Fan Lai}
\affiliation{%
  \institution{University of Illinois Urbana-Champaign}
  \city{Urbana}
  \state{Illinois}
  \country{USA}}
\email{fanlai@illinois.edu}

\author{Tianyin Xu}
\affiliation{%
  \institution{University of Illinois Urbana-Champaign}
  \city{Urbana}
  \state{Illinois}
  \country{USA}}
\email{tyxu@illinois.edu}

\author{Josep Torrellas}
\affiliation{%
  \institution{University of Illinois Urbana-Champaign}
  \city{Urbana}
  \state{Illinois}
  \country{USA}}
\email{torrella@illinois.edu}

% Running head on inner pages.
\renewcommand{\shortauthors}{Jaiswal et al.}

\begin{abstract}
    LLM agents alternate between inference and tool execution, repeatedly reusing their growing context across calls. Under high concurrency, their combined key-value (KV) cache working set can reach hundreds of GiB, far exceeding GPU memory capacity. As idle KV cache blocks are evicted to slower storage tiers, returning agents incur expensive retrievals that stall inference.

To address this problem, we present \emph{TripleKV}, a three-tier KV cache management system that hides the delays of these retrievals.
%exploits tool execution to hide these delays. 
In agentic workloads, the KV blocks needed by the next inference request are known in advance; only the time of their reuse is uncertain. TripleKV uses a tool-aware latency predictor to estimate when each tool call will complete and prefetches the required KV blocks from SSD into host DRAM ahead of time. Unlike GPU prefetching, this avoids reserving scarce GPU memory and interfering with concurrent inference requests. To further improve DRAM utilization, TripleKV eliminates duplication of GPU-resident blocks in DRAM, and introduces a victim-cache mechanism that refills blocks from SSD upon their eviction from GPU memory.

Across data, web, and coding agents, TripleKV reduces the fraction of requests requiring critical-path SSD reads from 45--76\% to under 10\%. It also reduces P95 time to first token (TTFT) by up to 49\% compared with vLLM extended with DRAM and SSD KV-cache tiers, and achieves  1.5-7$\times$ lower P95 TTFT than recent agent-aware systems.
\end{abstract}


\maketitle

\section{Introduction}
\begin{figure}[htbp!]
    \centering
    \includegraphics[width=0.75\columnwidth]{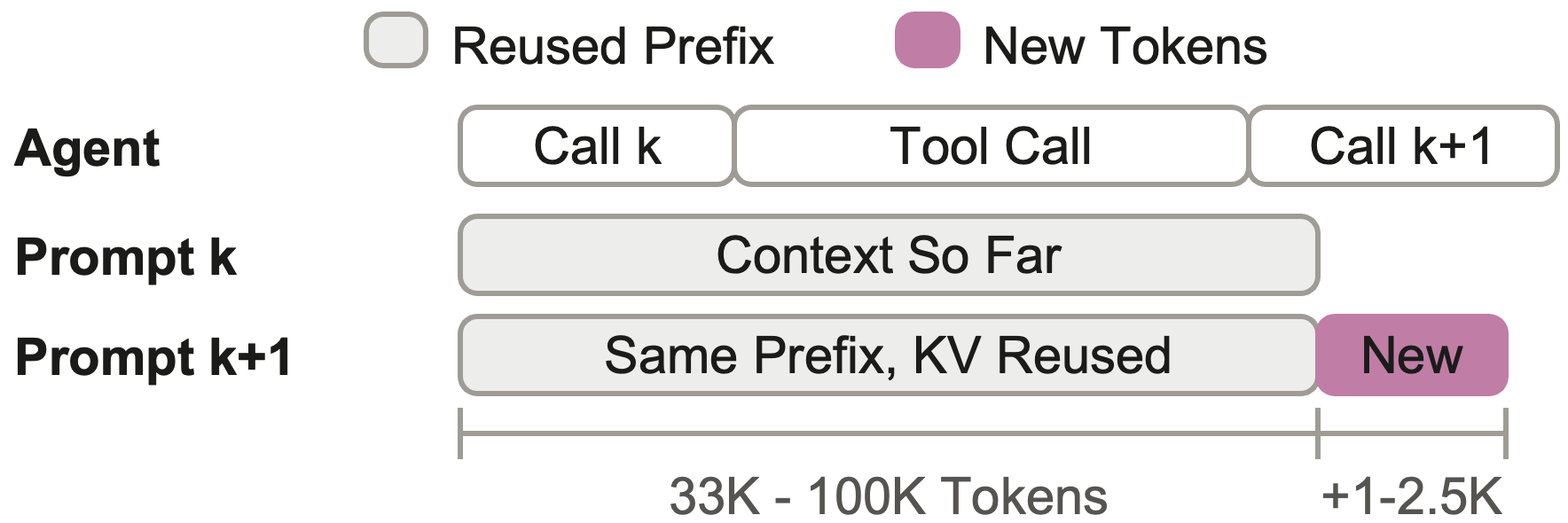}
    \caption{The ReAct agent loop.}
    \label{fig:react}
    \vspace{1mm}
\end{figure}

LLM agents have fundamentally changed the workloads handled by   inference servers, with ReAct (Reasoning and Action)~\cite{react} emerging as the de facto standard for AI agents solving complex tasks. ReAct-style agents alternate between LLM inference and tool calls (e.g., web searches, page visits, and shell commands), submitting their growing context as part of each new inference prompt (Fig.~\ref{fig:react}). As a result, agent contexts can grow to tens of thousands of tokens. To avoid recomputing these tokens, inference engines retain previously computed 
tokens in key-value (KV) caches, and reuse them across inference calls through prefix caching~\cite{vllm}. 
For Qwen3-30B-A3B~\cite{qwen30b}, a popular 30B-parameter mixture-of-experts model, the KV cache requires approximately 96 KiB per token. A single 50K-token agent context therefore occupies 4.6 GiB. On an NVIDIA H100 NVL, model weights and runtime overhead leave roughly 32 GiB for KV caching, enough for only about six such contexts.
Balancing KV cache reuse against GPU memory utilization is therefore critical to serving performance.

\begin{figure}[htbp!]
    \centering
    \includegraphics[width=\columnwidth]{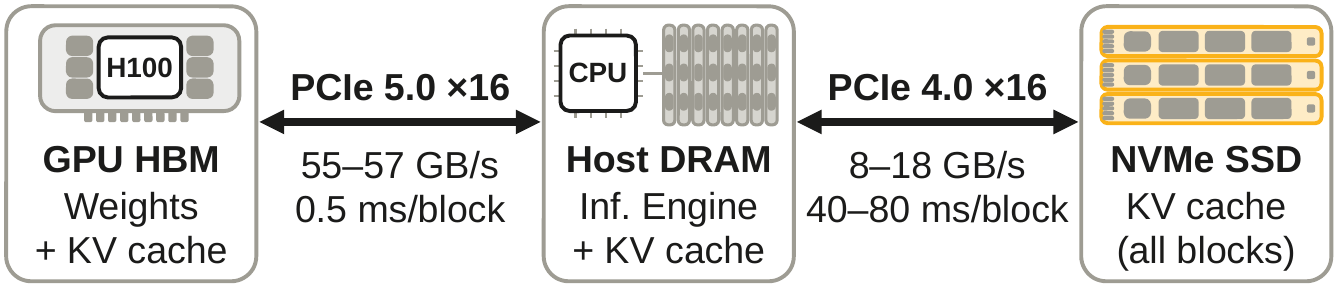}
    \caption{Memory hierarchy in LLM inference servers with measured KV cache transfer latency and bandwidth.}
    \label{fig:tiers}
\end{figure}

While an agent executes a tool call, its previously computed KV cache blocks may remain idle in GPU memory up to 100 seconds or more, awaiting the next inference request. Meanwhile, other agents compete for the same limited memory capacity, forcing serving engines to evict cached blocks, either discarding them or offloading them to host DRAM and potentially to slower storage tiers~\cite{lmcache,mooncake,dynamo} (Fig~\ref{fig:tiers}). 

Many KV cache management systems rely on Least Recently Used (LRU) policies for eviction. However, as we will show,
%in Figure~\ref{fig:block_lifespan}, 
LRU is poorly suited to agentic workloads even under moderate system load. Long tool execution intervals expose an agent's idle KV blocks to memory pressure from other agents' inference requests,  likely evicting them before the agent resumes inference. In our experiments, up to 76\% of inference requests find their previously cached blocks already evicted to slower tiers or discarded entirely. When the agent resumes inference, these blocks must either be retrieved from slower storage, potentially taking several seconds for a 50K-token context, or recomputed through an expensive prefill phase if discarded. Both introduce substantial delays, with recomputation additionally consuming GPU compute resources that could otherwise serve other agents.

Recent proposals explore different agent-aware strategies to manage KV caches.
Continuum~\cite{continuum} pins KV cache blocks in GPU memory throughout tool execution. This approach avoids reloads but occupies scarce GPU memory and limits throughput. TokenCake~\cite{tokencake} instead offloads KV blocks to host DRAM 
during tool execution. Later, it 
prefetches them back into GPU memory based on predicted tool latency. However, our observations show that prefetching into the GPU can be counterproductive: it competes with inference requests for GPU memory, increasing queueing delays.
Further, mispredictions waste precious GPU memory capacity on blocks that are not reused. Finally, while vLLM~\cite{vllm} is not agent-aware, it is worth mentioning
because it is   popular. vLLM supports SSD-backed KV caching to extend capacity beyond GPU and host memory. This avoids nearly all recomputation, as SSD is typically larger than the working set. However, retrieving evicted blocks from SSD can place storage I/O on the critical path of inference. Overall, existing approaches trade GPU memory capacity for reuse, risk contention and wasted GPU memory capacity through prefetching, or incur costly on-demand retrieval from slower tiers.

This paper approaches agent-aware KV cache management as a classical multi-level caching problem, where GPU HBM, host DRAM, and SSD are the 
three levels of the KV cache hierarchy. Our proposal is called {\em TripleKV}.
TripleKV relies on timely prefetching
of KV cache blocks from SSDs to host DRAM.
TripleKV includes three mechanisms: \textit{Predict}, \textit{Prefetch}, and \textit{Protect}. First,  a tool latency predictor estimates when each tool call will complete and schedules prefetching ahead of time to hide the expected storage read latency. Second, a prefetch engine transfers the agent's KV blocks from SSD to DRAM, using lower priority than read requests from inference calls. Finally,
a prefetch-aware cache replacement policy protects prefetched blocks in host
DRAM from premature eviction. 

TripleKV is built on top of vLLM~\cite{vllm}. 
It does not prefetch  directly 
into GPU memory because doing so introduces  contention and queueing delays.
Instead, when an agent resumes inference, vLLM's existing request admission mechanism retrieves the prefetched KV blocks from DRAM.
TripleKV relies on tool call information to timely prefetch  KV cache blocks before an agent's next inference request. Traditional hardware prefetchers must predict both what data will be accessed and when. In agentic workloads, the next data to access is known; what is not known is  when a tool call will
complete. Hence, TripleKV uses predictors. 
We find that tool calls are typically long
enough to hide SSD reads.

We evaluate TripleKV on four recent agentic benchmarks: DeepResearch Bench
(DRB)~\cite{du2025deepresearchbench}, GAIA~\cite{mialon2024gaia}, Terminal-Bench 4.0 (TB4)~\cite{merrill2026terminalbench}, and Browse Comp-Plus (BC+)~\cite{chen2025browsecompplus}. We run up to 20 concurrent agents on an NVIDIA H100 GPU. 
TripleKV reduces the fraction of requests requiring critical-path SSD reads 
from 45–76\% to
under 10\%. It also reduces P95 time to first token (TTFT) by up
to 49\% compared with vLLM extended with DRAM and SSD KV cache tiers, and
achieves 1.5-7$\times$ lower P95 TTFT than recent
agent-aware systems. Finally, it approaches the performance of an oracle with perfect knowledge of tool call completion times.

This paper makes four contributions:

\noindent $\bullet$ The TripleKV agent-aware KV cache management system that relies on predicting tool latencies and prefetching required KV cache blocks to host DRAM. 
 
\noindent $\bullet$ Characterization of KV cache behavior during agentic inference. %including different miss types, block lifespan, and other bottlenecks.

\noindent $\bullet$ KV cache block prefetching into host DRAM rather than
GPU memory, combined with   agent-aware cache eviction and replacement policy to prevent premature eviction.

\noindent $\bullet$ Evaluation of TripleKV, comparing to state of the art agent-aware baselines on latency, throughput, and  cache metrics.

\begin{comment}
\begin{enumerate}[leftmargin=*]
    \item Tool-latency predictors and a cache characterization of agents. A family of simple predictors, one per tool type, predicts a tool's end within 1s for 90\% tool calls.
    \item Prefetching KV cache data into L2 instead of L1. GPU memory caps the batch at a handful of contexts, so prefetching into it takes space from running requests and induces queuing. 
    \item A prefetching-aware cache eviction and replacement policy built using architectural concepts.
    \item TripleKV, an L1/L2/L3 cache model for KV cache memory that enables cheap external capacity off the critical path for storing KV cache without compromising the quality of service. 
\end{enumerate}
    
\end{comment}

\section{Background}

%\subsection{Agents Reuse Their Entire Previous Context Across Inference Calls}
\subsection{Agents Reuse Context at Every Step}

ReAct-style agents exhibit substantial context reuse across consecutive inference requests (Fig.~\ref{fig:react}). In all the representative benchmarks we study (Sections ~\ref{sec:motivation}, ~\ref{sec:methodology})  prompts can reach 100K tokens, and a single task can involve up to 100 LLM calls. Yet, each turn typically adds only 1.0K–2.5K new tokens. Thus, consecutive requests from the same agent share nearly their entire prefix, making KV cache reuse particularly important for efficient inference serving.

Serving engines exploit this reuse through prefix caching~\cite{prefix-caching}. Popular engines such as vLLM~\cite{vllm} and SGLang~\cite{sglang} store previously computed attention keys and values (KV cache) in fixed-size blocks, identified by hashes that incorporate both their contents and preceding prefix. When a new request arrives, the engine reuses matching blocks from the beginning of the prompt, computing only the remaining uncached suffix. In agentic workloads, 84.6–99.5\% of input tokens hit the prefix cache when capacity is sufficient~\cite{agentchar}, while studies of multi-turn chat report that up to 99\% of prefill work involves previously processed context~\cite{cachedattention}. In ReAct-style execution, each subsequent inference request follows the completion of a tool call, making tool completion the event that triggers reuse of the agent's existing context.

\begin{observation}
Agent inference requests repeatedly reuse large, contiguous KV cache prefixes. The blocks required by an agent's next request are largely known in advance, and their reuse is triggered by the completion of its current tool call
\end{observation}
\vspace{-3mm}

\subsection{The KV Hierarchy}
By default, vLLM stores KV cache blocks only in GPU memory. When a request completes, its blocks are placed on a free list but retain their contents and identities for future reuse. New allocations reclaim blocks from this list in Least Recently Used (LRU) order, effectively using available GPU memory as a cache of previously computed prefixes.

\begin{table}[t]
  \centering
  \caption{vLLM's tiered KV cache in cache terms.}
  \label{tab:cache-terms}
  \footnotesize
  \renewcommand{\arraystretch}{1.15}
  \begin{tabularx}{\columnwidth}{@{}>{\raggedright\arraybackslash}p{0.27\columnwidth}
                                     >{\raggedright\arraybackslash}X@{}}
    \toprule
    \textbf{Cache Term} & \textbf{In vLLM's tiered KV offload (our baseline)} \\
    \midrule
    L1                       & GPU HBM \\
    L2                       & CPU DRAM \\
    L3                       & SSD or any external storage \\
    Line                       & A KV cache block (order of MiBs)\\ %\xa{This is specific to your setup if you want to be generic replace to simply KV block (order of MBs)}\\
    Tag             & A hash of the token prefix\\
    Hit             & Block found in HBM (L1) or DRAM (L2) upon request admission \\
    Miss            & Block read from L3 \\
    MSHRs~\cite{kroft}         & The engine's queue of outstanding L3 reads \\
    Compulsory/cold miss     & Tokens never computed before, always prefilled \\
    Replacement                & LRU, separately in each tier \\
    Associativity              & Full: any block can take any slot (no conflict misses) \\
    \bottomrule
  \end{tabularx}
\end{table}

vLLM also supports tiered KV caching, extending GPU memory (L1) with host DRAM (L2) and external storage (L3). The hierarchy follows a write-through policy: completed KV blocks are copied to lower tiers while retaining their copies in faster memory. We use SSD as the external tier, with several terabytes of capacity, effectively eliminating L3 evictions. This three-tier hierarchy serves as our baseline (Table~\ref{tab:cache-terms}). Both GPU memory and DRAM use LRU replacement. When an inference request arrives, vLLM first checks for reusable blocks in GPU memory, retrieving missing blocks from DRAM or SSD as needed. SSD-resident blocks must first be loaded into DRAM before being transferred to GPU memory over PCIe. Although GPUDirect Storage~\cite{gds} can bypass this intermediate host-memory transfer, it cannot eliminate SSD read latency. 
%\xa{Broken citation here}

Unlike conventional processor caches, KV cache blocks are enormous. In our setup, each 64-token block occupies 6 MiB. Thus, retrieving a long prefix from SSD can take seconds. Moreover, prefix caching requires a contiguous sequence of available blocks; if a block is missing from all tiers, subsequent blocks cannot be reused as part of that prefix and must be recomputed. vLLM retrieves blocks on demand, only after the inference request arrives, without exploiting information about the agent's ongoing tool execution.

\begin{observation}
From the GPU's perspective, KV cache storage forms a three-level cache hierarchy, with GPU HBM as L1, host DRAM as L2, and SSD storage as L3.
\end{observation}

\subsection{\mbox{Agent-aware KV Cache Management Systems}}

In this subsection, we discuss existing agent-aware KV cache management systems. We also include vLLM as it is very popular.

\begin{figure}[t]
    \centering
    \vspace{2mm}
    \includegraphics[width=\columnwidth]{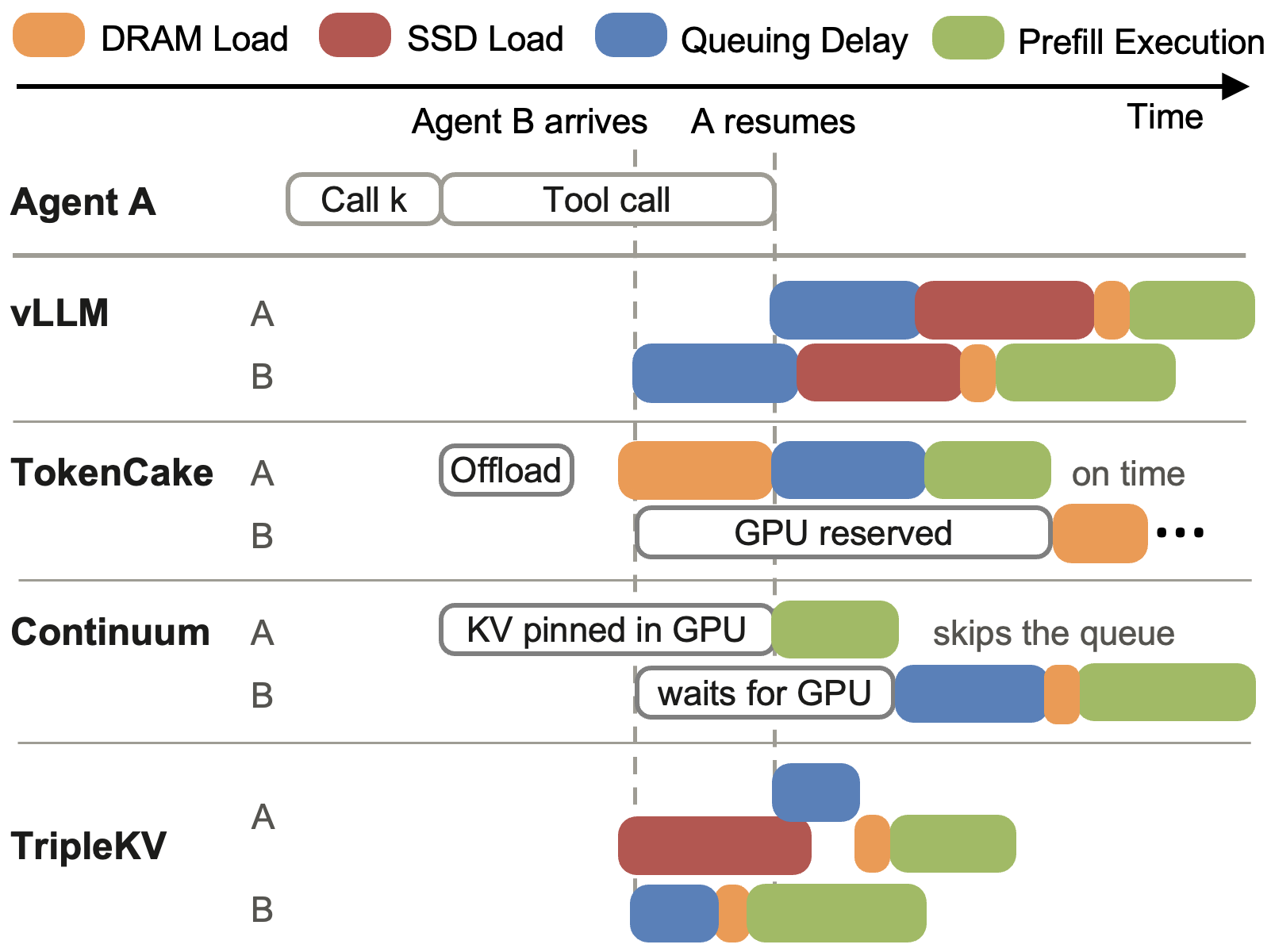}
    \caption{Timeline of one agent turn in various systems.}
    \label{fig:agent_turn}
    \vspace{3mm}
\end{figure}

While an agent executes a tool, its KV cache sits idle, and serving systems must decide where to keep it until the agent returns. Existing systems make one of three choices. vLLM~\cite{vllm} evicts the idle KV to DRAM or SSD and reloads it on demand once the next request arrives. TokenCake~\cite{tokencake} offloads it to DRAM and reloads it into GPU memory ahead of the agent’s predicted return. Continuum~\cite{continuum} keeps it pinned in GPU memory. Each choice trades GPU memory against reload latency in a different way.

Fig.~\ref{fig:agent_turn} illustrates these trade-offs for a single turn of an
agent A while a second agent, B, competes for the GPU.
Agent A issues call k and then executes a tool. Meanwhile, agent B arrives. In vLLM
(second row), A’s idle KV cache data is evicted during the tool call. Both requests first wait in the  GPU queue, and then read their KV cache context from the SSD into host DRAM, and from DRAM to GPU. Each pays the full SSD latency after admission. Because each request reserves GPU memory while its KV loads, the queue lengthens for every request behind it. 

TokenCake (third row) offloads A’s KV cache blocks to the DRAM and reloads them into the GPU before A’s predicted return. Note that reloading into the GPU takes a long time
because prefetches wait for empty GPU blocks and queue with other inference requests. A starts on time, but the GPU memory reserved for A blocks B until A finishes. At that point B can proceed.

Continuum (fourth row) pins A’s KV cache blocks in GPU memory for the whole tool call. A skips the queue when it resumes, but B waits for the pinned space to be freed before starting. The fifth row (TripleKV) will be explained  in Section~\ref{sec:overview}.

% TripleKV predicts when A’s tool will finish and prefetches A’s KV from SSD into DRAM during the tool call. The SSD read overlaps the tool call and A’s queueing delay, so after admission A pays only the short DRAM-to-GPU load. Since TripleKV neither reserves nor pins GPU memory, its queuing delays are shorter. Therefore, B is admitted after a shorter queuing delay, and both requests finish earlier than in other systems.
% \end{comment}
\section{Motivation}
\label{sec:motivation}
To characterize the performance of tiered KV cache offloading in agentic workloads, we deploy vLLM using the system setup in Table~\ref{tab:cache-terms} and the configuration in Table~\ref{tab:platform} (Section~\ref{sec:methodology}). We serve the Qwen3-30B-A3B model on an NVIDIA H100 NVL GPU, which has 32 GiB of GPU memory available for KV caching after accounting for model weights. We evaluate four benchmarks, DeepResearch Bench, GAIA, Terminal-Bench 4.0, and BrowseComp-Plus (Table~\ref{tab:workloads}), under increasing system load, varying the number of concurrently executing agents.

\subsection{Memory Limits the Number of Agents a GPU Serves}

\begin{figure}[t]
    \centering
    \includegraphics[width=0.9\columnwidth]{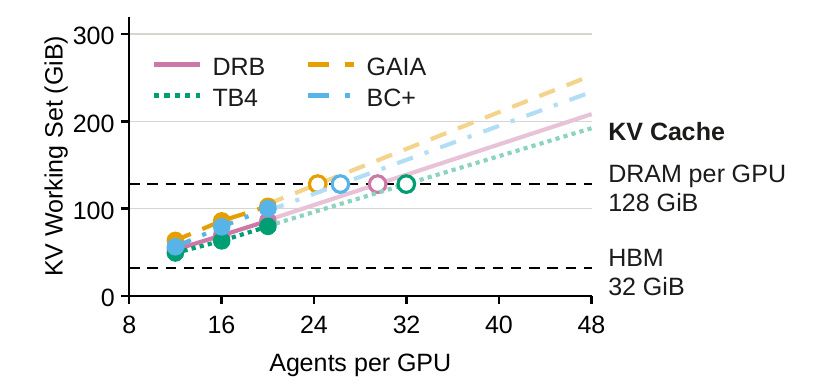}
    \caption{KV cache data of active agents per GPU.}
    \label{fig:kv_per_gpu}
    \vspace{1mm}
\end{figure}

Figure~\ref{fig:kv_per_gpu} shows how the aggregate KV cache working set grows with agent concurrency. We define the working set as the total KV cache state of all agents with unfinished tasks, including those waiting for tool calls. At 20 concurrent agents, it reaches 127 GiB, nearly $4\times$ the 32 GiB of GPU HBM available for KV caching. With each agent contributing up to 8 GiB, the working set grows approximately linearly with concurrency. Because DRAM serves as an inclusive cache, retaining the entire working set without accessing SSD requires sufficient DRAM capacity to accommodate all KV blocks, including those also resident in GPU memory. The horizontal dotted line marks a reference capacity of 128 GiB of host DRAM per GPU, illustrating how quickly increasing agent concurrency can exhaust this capacity.

\begin{figure}[t]
    \centering
    \includegraphics[width=0.9\columnwidth]{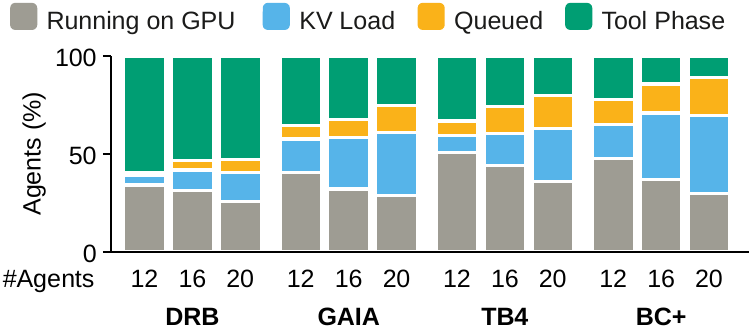}
    \caption{Percentage of agents in different phases of execution on average.}
    \label{fig:agent_time}
\end{figure}

\noindent Figure~\ref{fig:agent_time} shows the fraction of agents in different execution phases as we increase agent concurrency. Regardless of the number of active agents, the GPU executes only 4–9 inference requests concurrently, largely constrained by the number of contexts that fit in its HBM. The remaining agents must wait their turn, often requiring their evicted KV cache blocks to be restored from DRAM or slower tiers before resuming inference. At high concurrency, up to 14 agents are waiting to execute, with 54-74\% of them waiting for their KV cache data to be loaded. Overall, up to 50\% of active agents, on average, are stalled waiting for KV cache restoration in the vLLM baseline. These results highlight that, as the aggregate KV cache working set exceeds GPU memory capacity, efficiently moving KV blocks between memory tiers becomes critical to serving performance.

One approach to avoid KV cache restoration is to retain idle
contexts in GPU memory. Continuum~\cite{continuum} follows
this approach, pinning an agent's KV cache for a time-to-live
(TTL) estimated from tool latencies. While this avoids
reloading the context when the agent returns, it reduces
GPU memory available to other requests. We evaluate this
tradeoff further in Section~\ref{sec:eval}.
\vspace{2mm}

% \added{In our evaluations in Section \ref{sec:eval}, we give the system 64 GiB of DRAM for KV cache, for two reasons. First, a finished session's KV stays useful for the user's next turn or a follow-up task and stays on the DRAM as well. In our experiments, a 50 agent session leaves 200–500 GiB of context, which is more than the available host memory. Second, the working set scales with the model. A comparable dense model, Qwen3-32B\cite{qwen3-32b}, generates $2.7\times$ the KV per token and needs $2.7\times$ the DRAM. Moreover, the ratio of the working set to the amount of DRAM available is what decides how the system behaves. At 12–20 agents, 64 GiB holds $0.5-1.4\times$ W, the regime in which that ratio reaches and exceeds one.}
% \xa{This is not the place for this -- you should move this to methodology. The first part I do not get. The second part I get it a bit, you could write something like: We limited host memory to 64GB to keep the observations representative for larger models and higher load? Keep it short. If you get the chance ask also Josep.}

\begin{observation}
GPU memory capacity limits the number of concurrently executing inference requests, while the aggregate KV cache working set far exceeds available HBM. Consequently, KV cache restoration becomes a major source of agent waiting time, limiting how quickly the GPU can serve new requests.
\end{observation}

\subsection{L3 Reads in Critical Path}

\begin{figure}[t]
    \centering
    \includegraphics[width=0.9\columnwidth]{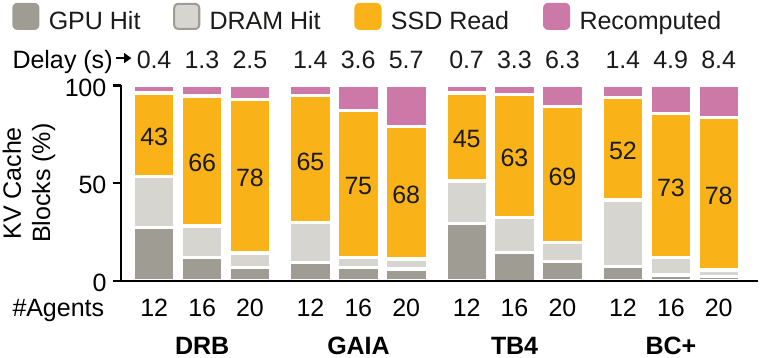}
    \caption{Source of each prompt token's KV in vLLM with external tier.}
    \label{fig:kv_source}
    \vspace{2mm}
\end{figure}

Fig.~\ref{fig:kv_source} shows the average distribution of requested KV cache blocks (per prompt) across cache tiers in our vLLM baseline. Only 2–33\% of tokens are served from GPU memory and 0–28\% from host DRAM, while up to 78\% must be retrieved from SSD after request admission. In addition to the results shown in the figure, we observe that up to 76\% of inference requests stall waiting for SSD reads before generating their first token. These delays are substantial; retrieving a 5 GiB context from SSD takes approximately 1–2.5 seconds in our setup under average load. The numbers above the bars in Fig.~\ref{fig:kv_source} report the average time each request spends waiting for its KV cache blocks to be loaded.

The figure also shows that the fraction of recomputed KV blocks increases with the number of agents. As concurrency grows, more requests transfer KV cache data between SSD and GPU, increasing pressure on host DRAM as well as the interconnect bandwidth. In vLLM's inclusive cache hierarchy, DRAM must retain copies of GPU-resident blocks while also accommodating in-flight transfers to and from SSD. Under high contention, insufficient DRAM capacity can prevent some transfers from completing. When a retrieval exceeds vLLM's timeout, the engine abandons the transfer and recomputes the missing KV blocks instead. Consequently, higher load can lead to more recomputation, even when the required KV blocks remain available on SSD.

\begin{figure}[htbp!]
    \centering
    \includegraphics[width=0.9\columnwidth]{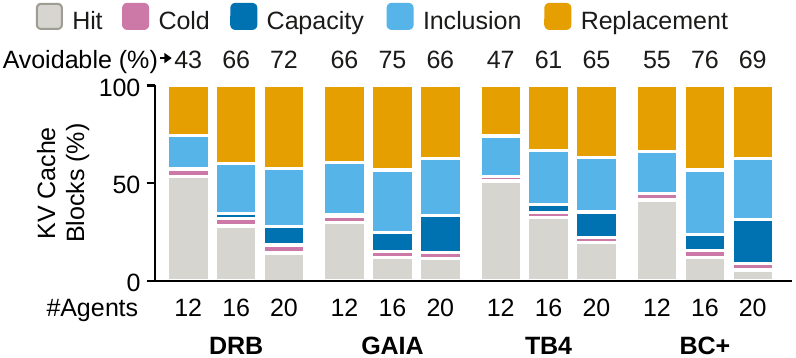}
    \caption{Hits and Misses.}
    \label{fig:miss_types}
    \vspace{2mm}
\end{figure}

\noindent Fig.~\ref{fig:miss_types} breaks down KV cache misses by their underlying cause. For this analysis, we define a cache hit as finding the required KV block in either HBM or DRAM, and a miss as requiring access to SSD or recomputation. Following classical cache replacement-policy studies~\cite{hawkeye,mockingjay}, we compare LRU against Belady's optimal replacement policy, MIN~\cite{Belady}, and distinguish four types of misses:

\begin{itemize}[leftmargin=*]
    \item \textbf{Cold Miss:} The required KV block has never been computed.
    \item \textbf{Capacity Miss:} Even MIN misses when HBM and DRAM are treated as a single exclusive cache, indicating that their combined capacity is insufficient.
    \item \textbf{Inclusion Miss:} MIN hits with an exclusive cache hierarchy but misses with an inclusive DRAM tier, as duplicate copies reduce the effective cache capacity.
    \item \textbf{Replacement Miss:} LRU misses while MIN hits under the same inclusive hierarchy, indicating a suboptimal eviction decision.
\end{itemize}

\begin{figure}[t]
    \centering
    \includegraphics[width=\columnwidth]{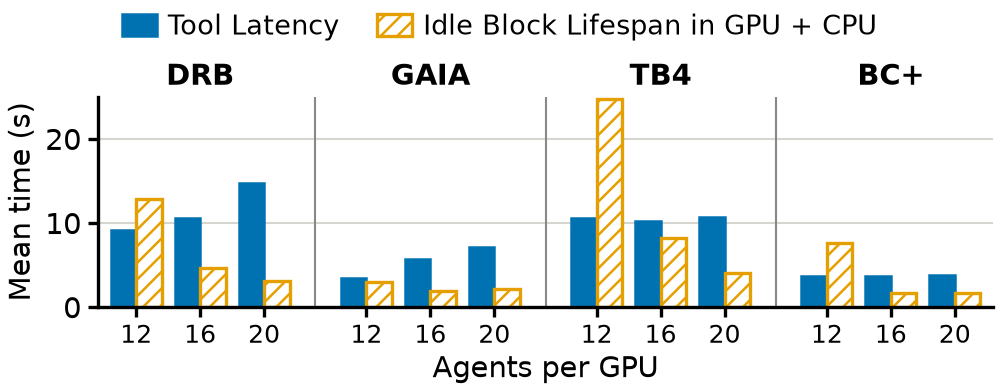}
    \caption{An idle block's lifespan under LRU in L1 and L2 compared to tool latencies.}
    \label{fig:block_lifespan}
    \vspace{2mm}
\end{figure}

\noindent We observe that avoidable inclusion and replacement misses dominate, with their combined percentages shown above each bar in Fig.~\ref{fig:miss_types}. To further investigate replacement misses, we measure the average time an idle KV block remains in the L1 and L2 tiers under LRU. Fig.~\ref{fig:block_lifespan} shows that tool execution times are typically much longer than this cache residency time, causing KV blocks to be evicted before their agents resume inference. As other agents compete for cache capacity during tool execution, LRU evicts the idle agent's blocks before they can be reused.

\begin{observation}
Most KV cache misses are avoidable. LRU is suboptimal for agentic workloads because tool execution times are often longer than the time KV blocks remain cached, leading to eviction before reuse.
\end{observation}

\subsection{What, When and Where to Prefetch?}
\label{sec:obs_prefetch}

The above findings suggest that retaining every agent's KV cache in fast memory is impractical, but retrieving evicted blocks only after the next inference request arrives is expensive. Instead, we investigate whether tool execution provides an opportunity to bring the required KV blocks back before they are needed. This raises three questions: what, when, and where to prefetch?

\begin{figure}[h!]
    \centering
    \includegraphics[width=0.9\columnwidth]{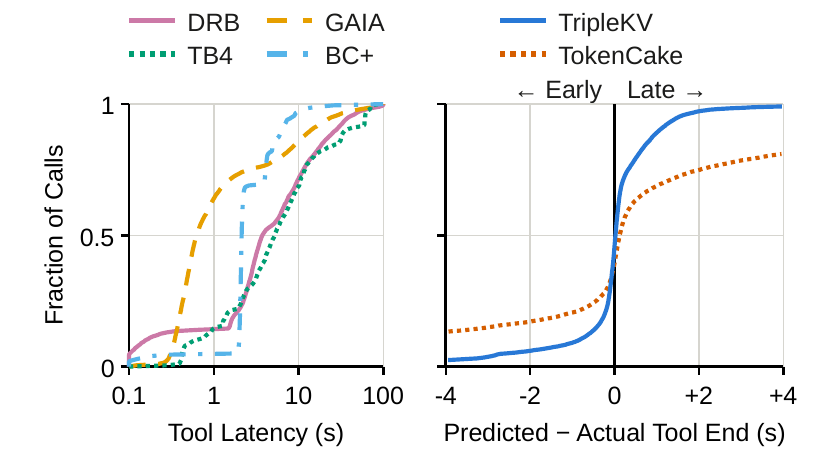}
    \caption{Tool Latency (left) and Prediction Error (right).}
    \label{fig:tool_latency}
    \vspace{2mm}
\end{figure}

\paragraph{What to Prefetch?}
Traditional hardware prefetchers must predict both what data to fetch and when. For ReAct-style agents, the first question is largely resolved: an agent's next inference request reuses its existing KV cache prefix, which is known in advance. Only the timing remains uncertain, as reuse follows the completion of the tool call. This resembles informed prefetching~\cite{patterson}, where applications disclose future data accesses.

\paragraph{When to Prefetch?}
Fig.~\ref{fig:tool_latency} (left) shows that tool execution times provide sufficient time windows for prefetching. Median tool latencies reach up to 10 seconds, while retrieving a 5 GiB context from SSD at 2–5 GB/s takes approximately 1–2.5 seconds. Accurately predicting tool completion times is though important, as prefetching too early wastes cache capacity, while prefetching too late leaves data retrieval on the inference request's critical path. Concurrent work, TokenCake~\cite{tokencake}, uses moving averages of past tool latencies to predict completion times. As shown in Fig.~\ref{fig:tool_latency} (right), fewer than 50\% of TokenCake's predictions fall within $\pm 1$ second of actual completion. However, we find that incorporating tool-specific characteristics, such as tool inputs, and tailoring predictors to individual tools improves this accuracy to over 90\% of calls. We describe these predictors in Section~\ref{sec:predict}. Even imperfect predictions can be useful, as partial prefetching can reduce the several-second KV-loading delays observed under high concurrency (Fig.~\ref{fig:kv_source}).

\paragraph{Where to Prefetch?}
TokenCake~\cite{tokencake} offloads idle KV cache blocks to
host DRAM and prefetches them back into GPU memory ahead
of predicted tool completion. This hides the DRAM-to-GPU
transfer but requires scarce HBM capacity before the request
can execute. With GPU memory accommodating only 5--7
average agent contexts, prefetches compete with inference
requests for free GPU memory, increasing queueing delays.
We evaluate this effect in Section~\ref{sec:eval}.
Moreover, SSD-to-DRAM transfers are significantly slower
than DRAM-to-GPU transfers (Fig.~\ref{fig:tiers}).
This suggests that prefetching into DRAM can hide the
dominant transfer cost without reserving GPU memory
in advance.

%\begin{observation}
%An agent's reusable KV cache blocks are known in advance, and (predicted) tool completion times provide a useful signal for when they will be needed. Prefetching into DRAM can hide the dominant storage latency without competing with inference requests for GPU memory.
%\end{observation}

\begin{observation}
An agent's reusable KV cache blocks are known in advance, and tool completion times can be accurately predicted to determine when they will be needed. Prefetching into DRAM hides the dominant storage latency without competing with inference requests for scarce GPU memory.
\end{observation}

\section{TripleKV Overview}
\label{sec:overview}
\begin{figure*}[htbp!]
    \centering
    \includegraphics[width=0.9\textwidth]{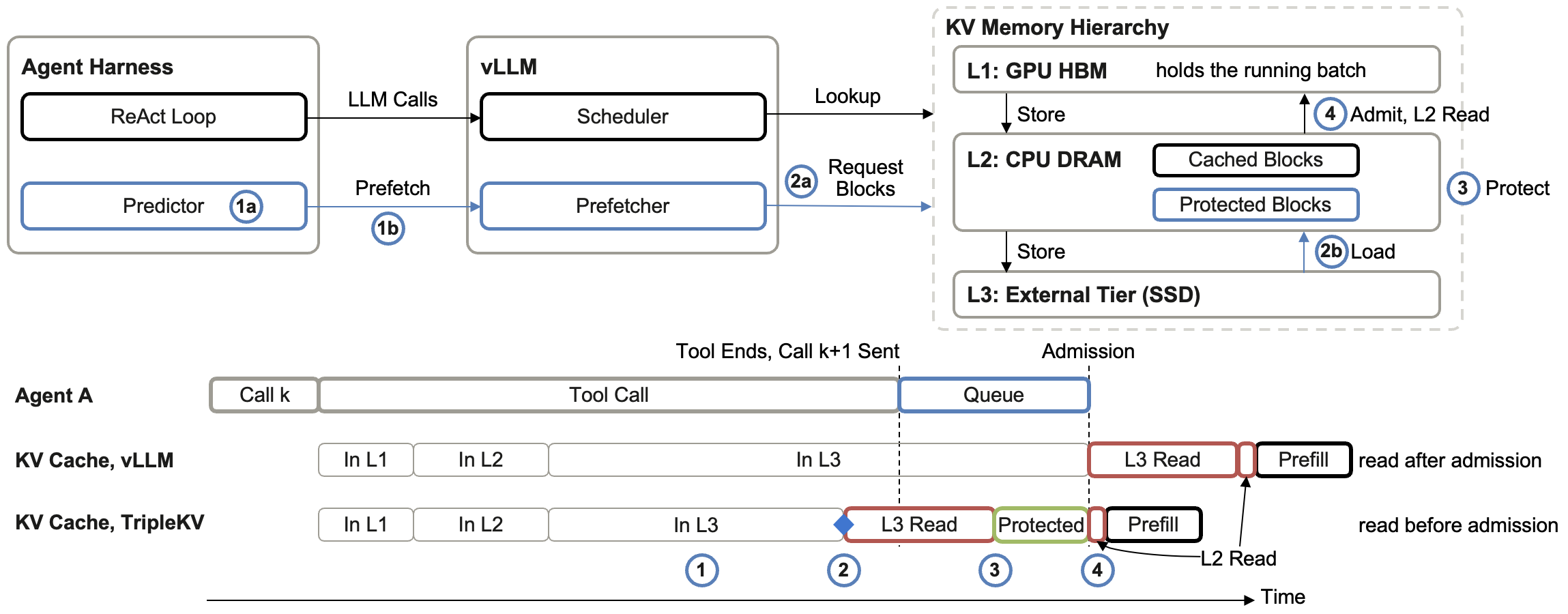}
    \caption{TripleKV Control Flow (top) and KV Cache Data Movement (bottom).}
    \label{fig:arch}
    \vspace{2mm}
\end{figure*}

Guided by the observations in Section 2, TripleKV introduces a KV cache management system that uses prefetching to hide the latency of data movement across memory tiers. TripleKV builds on vLLM's three-level KV cache hierarchy, with GPU HBM, host DRAM, and SSD serving as L1, L2, and L3, respectively (Observation 2). \added{It breaks the inclusive property of the L2 tier in vLLM and makes it a non-inclusive tier to fit more blocks in faster tiers. L3 remains inclusive and every block that is generated in the GPU is written to it asynchronously.}
It preserves vLLM's existing demand path, where KV blocks are loaded into GPU memory only after request admission, while extending L2 cache management to support prefetching. 
TripleKV introduces three mechanisms:

%Section~\ref{sec:motivation} showed that during tool execution, an agent's KV cache blocks are often evicted from GPU memory and DRAM, leaving their copies in the external SSD tier. vLLM retrieves these blocks only after the agent's next inference request is admitted, placing expensive SSD reads on the request's critical path (Observation 3). However, the blocks the agent will reuse are known in advance, and tool completion times can be predicted (Observation 5). TripleKV exploits this opportunity to prefetch KV blocks from SSD into DRAM during tool execution and retain them until the agent resumes inference.

\begin{itemize}[leftmargin=*]
    \item \textbf{Predict:} A tool latency predictor in the agent harness estimates when each tool call will complete and schedules prefetching ahead of that time.
    \item \textbf{Prefetch:} A prefetch engine in vLLM asynchronously retrieves the agent's KV blocks from L3 into L2.
    \item \textbf{Protect:} A prefetch-aware cache eviction and replacement policy retains prefetched blocks until they are reused to avoid premature eviction.
\end{itemize}

TripleKV makes two key design choices. First, it never prefetches into L1, avoiding contention for scarce GPU memory (Observation 3 and 5). Instead, prefetched blocks remain in L2 until the request is admitted, when vLLM transfers them into L1 over PCIe. This transfer remains on the critical path but is relatively inexpensive ($\approx 200$ ms). Second, TripleKV does not predict which KV blocks will be accessed. The agent's existing context identifies the required blocks (Observation 1), which are located using the same token-based hashes as vLLM's prefix caching. Thus, only the timing of the prefetch must be predicted (Observation 5). 
% \xa{How about the inclusiveness thing?}

Fig.~\ref{fig:arch} illustrates TripleKV's operation during one turn of agent A in detail. While A executes a tool call, competing requests may evict its KV blocks from L1 and L2, leaving their copies in L3. The predictor estimates when the tool will complete \circled{1a} and schedules a prefetch request containing the agent's context \circled{1b}. The prefetch engine then initiates the required reads \circled{2a}, transferring A's blocks from L3 into L2 \circled{2b}. These blocks remain protected from eviction until they are reused or their protection expires \circled{3}. When A's next request is admitted, vLLM retrieves the blocks from L2 into L1 and prefills only the new tokens from the prompt \circled{4}. 
% \xa{but you also fetch from disk and recompute when things are late} 
\added{If the prefetch does not finish before the request arrives, the request stalls for only the remaining transfer. However, this happens rarely in our experiments, for less than 10\% of the requests even under high load.}
In contrast, baseline vLLM initiates the SSD reads only after request admission, forcing the request to wait for their completion (Fig.~\ref{fig:arch}, bottom).

TripleKV is more efficient than existing agent-aware KV cache management systems.
Going back to 
Fig.~\ref{fig:agent_turn} (fifth row), we see how both agents A and B finish their
requests faster. TripleKV predicts when A’s tool will finish and prefetches A’s KV cache blocks from SSD into DRAM during the tool call. The SSD read overlaps the tool call and A’s queueing delay. As a result, after admission, A pays only the short DRAM-to-GPU load. Since TripleKV neither reserves nor pins GPU memory, its queuing delays are shorter. Therefore, B is admitted after a shorter queuing delay.%, and both requests finish earlier than in other systems.

\textbf{Protection in L2.}
TripleKV modifies vLLM's cache replacement and eviction policy for the L2 tier to prevent prefetched blocks from being prematurely evicted (Observation 4). Instead of using naive LRU, we manage L2 using a multi-queue~\cite{zhou2001mq} replacement policy that uses three queues with different eviction priorities. Prefetched blocks enter the queue with the lowest eviction priority and stay protected until their request is admitted to HBM (or until a TTL expires). 
% If a block that is going to be reused already has a copy in L1, it is kept in the backup queue. Every other block remains in the cached queue. Each queue evicts its blocks in LRU order but the prefetched queue has lowest eviction priority, followed by the backup queue. 
We explain this in detail in Section \ref{sec:policy}.

When L2 needs space, the ordinary cached block queue evicts first, in LRU order.
% \xa{The problem here is that we spend so much time saying LRU is bad but we end up using it -- it is just that you do not sue it on the HBM? What is the special eviction policy we talk about on intro? Is it the protection mainly? If you have the time I think it makes sense to try and present it better here.}. 
Protected blocks are evicted only when necessary, starting with those whose protection expires soonest. Protection ends when a block is reused or its timeout expires, limiting the impact of inaccurate predictions. To preserve normal cache operations, protected blocks remain evictable under memory pressure. Thus, prefetching cannot prevent critical path reads or writes from completing. %We describe the replacement policy in Section~\ref{sec:policy}.
\section{Design}

TripleKV has three phases: predict, prefetch and protect.%: a predictor times each prefetch. Prefetch: an engine in vLLM carries out the prefetch. Protect: L2's replacement order protects the prefetched blocks. Blocks that L1 evicts are refilled, and SSD queues keep prefetches out of the way of demand misses.

\subsection{Predict: When Will the Tool Return?}
\label{sec:predict}
\begin{figure}[htbp!]
    \centering
    \includegraphics[width=0.8\columnwidth]{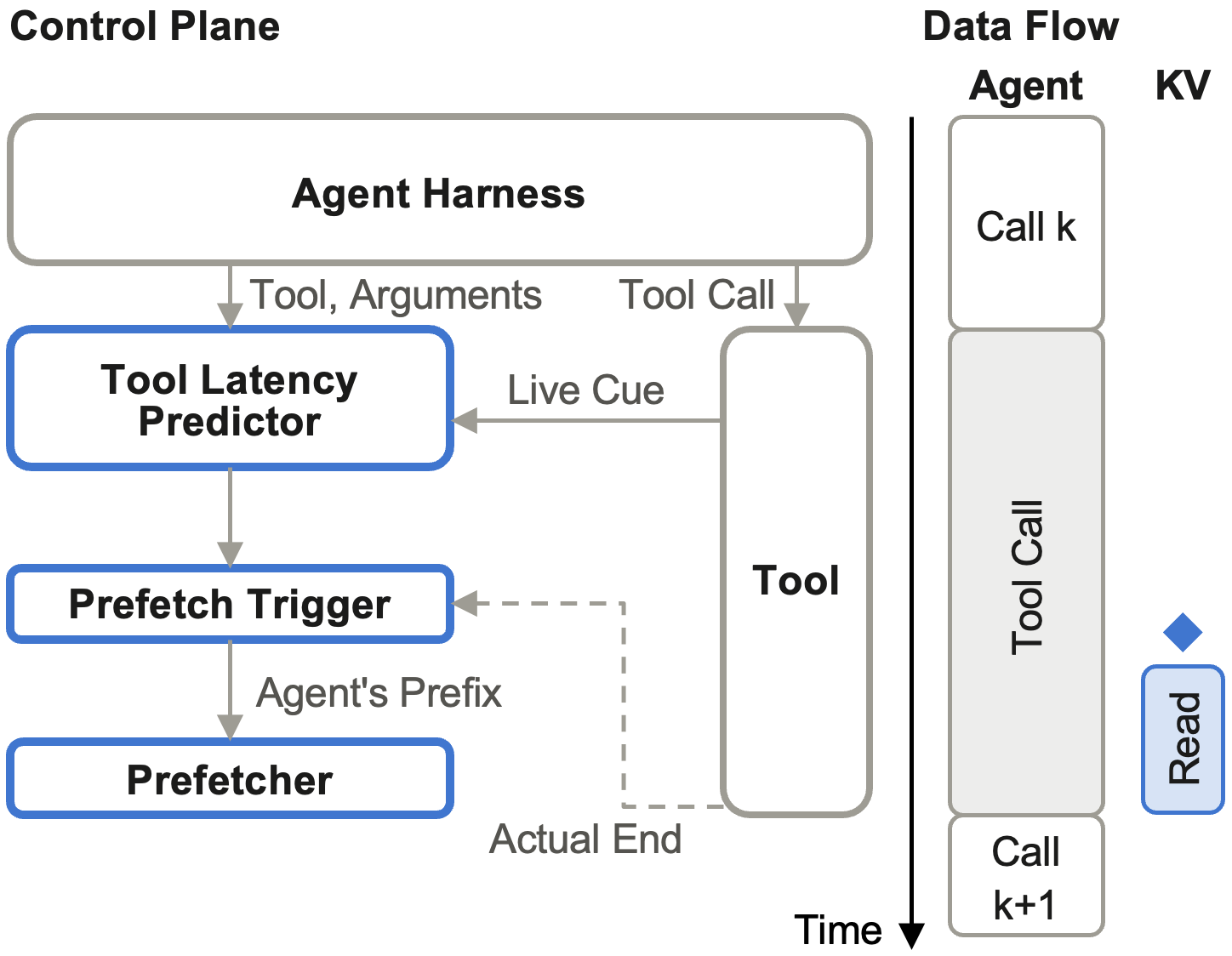}
    % \newline 
    \caption{The Tool Latency Predictor.}
    \label{fig:prefetch}
\end{figure}

The tool latency predictor (Fig \ref{fig:prefetch}) predicts the time when the tool call is expected to end. It runs in the agent harness, which sees each tool call start, its arguments, and its output as it arrives. Like a branch predictor that keeps a history per branch, it keeps a short history per tool, including the latencies of that tool's recently finished invocations. Prior works \cite{tokencake, continuum} also maintain estimates per tool, but they predict from durations alone. TripleKV's estimates also read the call's input arguments and its live output (if available), and their form differs between different kinds of tool types. The top ten most popular tools and their predictors are described in Table~\ref{tab:tool-predictors}. Across all 18.5K tool calls, TripleKV’s predictors have an MAE (Mean Absolute Error) of 1.1 s versus 7.6 s for TokenCake’s predictors, and a MAPE (Mean Absolute Percentage Error) of 21\% versus 117\%. Fig \ref{fig:tool_latency} (right) visualizes this error distribution. %90\% of TripleKV's tools are predicted within $\pm1s$ of their actual latency as opposed to only 50\% for TokenCake.

% needs \usepackage{booktabs,tabularx}
\begin{table}[htbp!]
\centering
\caption{The top 10 most popular tools and their latency prediction technique in TripleKV.}
\label{tab:tool-predictors}
\footnotesize
\setlength{\tabcolsep}{3pt}
\renewcommand{\arraystretch}{1.1}
\begin{tabularx}{\columnwidth}{@{}>{\raggedright\arraybackslash}p{0.25\columnwidth}
                                >{\raggedright\arraybackslash}p{0.27\columnwidth}
                                >{\raggedright\arraybackslash}X@{}}
\toprule
\textbf{Tool type} & \textbf{Signal} & \textbf{Predicted tool time} \\
\midrule
Web search
  & Number of queries in the call
  & Queries $\times$ median per-query latency, plus median overhead \\
Page fetch
  & Payload size
  & Linear predictor plus median overhead \\
Page summary (LLM)
  & Live cue: the last page's summary begins
  & That moment, plus a conservative estimate of the time left \\
Scholarly search
  & Tool name
  & Median recent latency of the tool \\
Code interpreter
  & Tool name
  & Median recent latency of the tool \\
File parser
  & Tool name
  & Median recent latency of the tool \\
% Turn with no tool call
%   & No tool in the reply
%   & Median recent length of such turns \\
% \midrule
% \midrule
Terminal command
  & Wait time the agent requests
  & Requested wait, plus this task’s recent extra delay \\
Terminal turn without a command
  & None
  & This task’s recent delay (no wait requested) \\
New terminal task
  & No history for this task yet
  & Requested wait, plus other tasks' usual extra delay \\
Python interpreter & Command name & Median recent latency of the command \\
% \midrule
% Any tool, no history yet
  % & No finished call of the tool
  % & None; the prefetch is issued at the tool's start \\
\bottomrule
\end{tabularx}
\end{table}

The prefetch is scheduled to be issued with just enough time left to load the blocks before the request is to be scheduled as described in Equation~\ref{eq:issue}. %at the earlier of the predicted and the actual end. 
If the tool returns before the prefetch was issued, the harness recalculates the prefetch issue time based on the observed tool latency. A prefetch that is issued early keeps its blocks protected in L2 for longer. One that is issued late leaves part of the KV read on the request's critical path.

Fig \ref{fig:prefetch} illustrates the tool latency predictor. The control plane is shown on the left, and the data flow of one tool call is shown on the right, on a single time axis running downward. It predicts each tool's end from the tool type, its arguments, recent latencies and live cues. The prefetch is issued before the tool ends, such that the KV cache loading finishes before request k+1 is admitted to be scheduled and may start without blocking. 

With an estimate of when the tool ends, the best time to issue
the prefetch is the latest instant that still completes the read by the time the request is admitted:
\begin{equation}
  t_{\mathrm{issue}} = t_{\mathrm{end}} + q - R - \epsilon,
  \qquad R = \frac{S_{3}}{B},
  \label{eq:issue}
\end{equation}
where $t_{\mathrm{end}}$ is the time the tool ends, $q$ is the current queueing
delay of the system (read live from vLLM's counters), $R$ is the time to
read the part of the agent's context held only in L3, $S_{L3}$ is the
size of that part, $B$ is the SSD bandwidth available to prefetches, and
$\epsilon$ is a small constant lead for safety. Issuing a prefetch earlier than $t_{\mathrm{issue}}$ only keeps blocks
protected for longer. %Whenever $q \geq R + \epsilon$, $t_{\mathrm{issue}}$ falls after the tool ends, so firing at the tool's end is already optimal.

In TripleKV, we observe that only less than 9\% of the KV cache data read from L3 lies on the critical path of the request on average. In contrast, with vLLM it is up to 74\%. On the other hand, early prefetching is also rare in TripleKV and costs little. In our experiments, less than 8\% of prefetches finish more than 1\,s before the request is admitted.
% \hf{do we observe how many prefetch requests have completed but not been used yet. That gives a bound on the L2 memory used for prefetched requests.}

% The dashed arrow is the fallback: if the tool ends first, its actual end fires the prefetch. TripleKV fires at min(predicted, actual), with no lead (§4.2). Early prefetches hold DRAM longer; late ones expose part of the read.

% Each kind of tool gets the simplest estimate that fits it:

% Fig \ref{fig:prefetch}

\vspace{-2mm}
\subsection{Prefetch: Moving Data from L3 to L2}
The harness sends a prefetch as one API call to vLLM, with the token IDs of the agent's context and a timeout. vLLM hashes the tokens into 64-token block keys exactly as it would for a prompt, and the prefetch engine walks the blocks in order. The engine runs inside vLLM's scheduler loop. Between engine steps, it claims the blocks already in L2, and then performs lookups and reads in L3 which go through the same path that serves requests, at prefetch priority. 
%A prefetched block is therefore an ordinary L2 block, and admission finds it with the usual lookup. 
A prefetch stops at the first block missing from every tier, since the request must recompute everything after it anyway.%, or when protecting more blocks would cut into the memory reserved for in-flight blocks.

\begin{figure}[t]
    \centering
    \includegraphics[width=0.85\columnwidth]{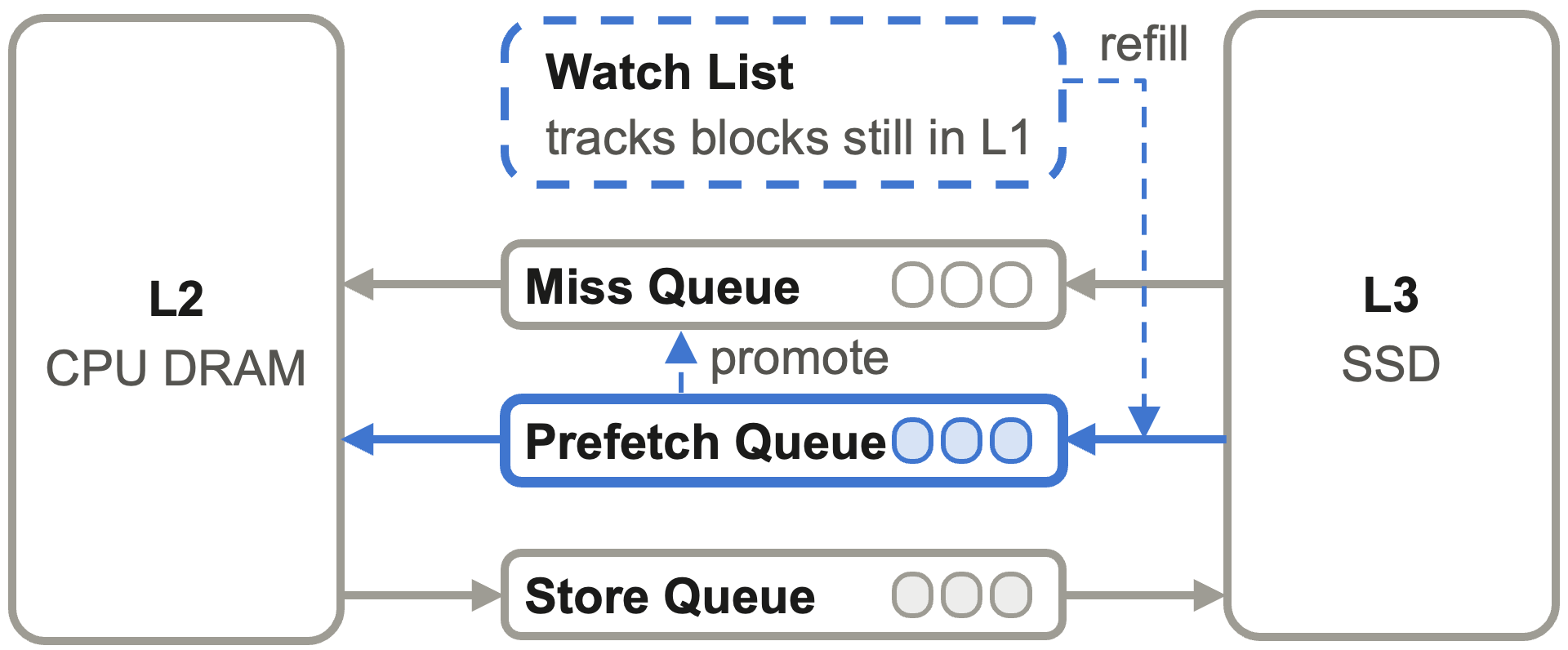}
    \caption{Read and Write Queues between L2 and L3.}
    \label{fig:queues}
     \vspace{1mm}
\end{figure}

Every L3 access goes through one of the three FIFO queues as shown in Fig \ref{fig:queues}. The miss queue holds reads that are in the critical path of a waiting request. The prefetch queue holds prefetch and refill reads.
A refill also fetches a block from L3 into L2. It is issued when L1 evicts a watched block. %A block is added to the watch list if it was present in L1 but not in L2 when its prefetch was issued. 
When a block to be prefetched is found to be in L1, TripleKV adds its hash to a watch list instead of prefetching it into L2, to avoid redundancy. If the block in L1 is evicted before the request can use it, the watch list adds it to the prefetch queue as a refill read so that it can be loaded into L2. This mechanism behaves like a victim cache~\cite{jouppi1990victim}.
% This is needed as TripleKV needs to maintain the entire context of the request within L1 and L2.

The store queue holds the write-through stores of new blocks. Each 6 MiB block is one data movement task, so the reads of one context spread over many threads. Two groups of threads drain the queues in different orders. The read-priority threads take misses first, then prefetches, then stores, so no thread starts a prefetch read while a miss is waiting. The write-priority threads take stores first, then misses, then prefetches, so the write-through keeps moving under heavy read load and neither direction starves. When a request arrives while its own prefetch is still queued, those tasks are promoted to the miss queue and are read as misses.

\subsection{Protect:  Eviction and Replacement  for L2}
\label{sec:policy}
A prefetch helps only if its blocks are still in L2 when the request arrives. %Every prefetched block also claims a victim to evict\cite{cao}. 
In order to ensure this, TripleKV uses a prefetch-aware cache eviction and replacement policy for L2. This is inspired from a multi-queue replacement policy \cite{zhou2001mq} that uses three queues with different eviction priorities. 
\begin{figure}[t]
    \centering
    \includegraphics[width=0.9\columnwidth]{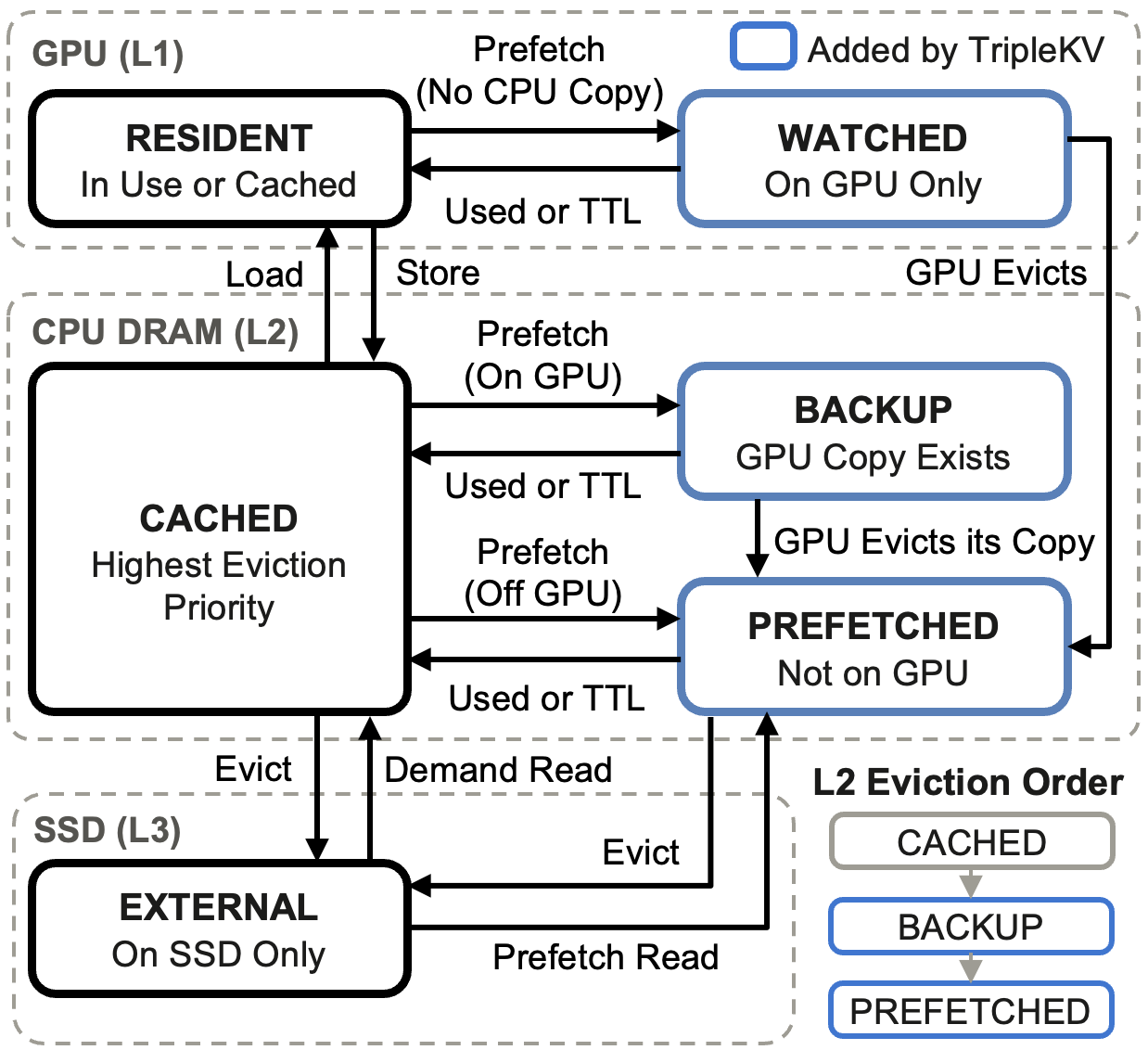}
    \caption{KV Cache Block State Diagram.}
    \label{fig:block}
    \vspace{2mm}
\end{figure}

Each L2 block lives in one of the following three classes (and their corresponding LRU queues). Fig~\ref{fig:block} shows the state diagram.
\begin{itemize}[leftmargin=*]
    \item PREFETCHED: A prefetched block that L1 does not hold.
    \item BACKUP: An L2 copy of a block that L1 still holds.
    \item CACHED: All other blocks.
\end{itemize}

The eviction priority is CACHED > BACKUP > PREFETCHED. Within each queue, LRU is followed with entering the queue counting as a use. This is done to ensure that PREFETCHED blocks are evicted as the oldest prefetched first (same for BACKUP). A BACKUP is promoted to a PREFETCHED block when L1 evicts its copy, because L2 then holds the only fast copy.
% In the PREFETCHED queue, the time these blocks were added to the queue is also considerd a use and is used when determining LRU. The same is done for the BACKUP queue.

% When L2 needs space, it evicts CACHED blocks first, in LRU order. BACKUP blocks are evicted next, oldest first, since L1 still has a copy. PREFETCHED blocks go last, earliest expiry first. Blocks in transfer are never evicted. A BACKUP is promoted to a PREFETCHED block when L1 evicts its copy, because L2 then holds the only fast copy.

Protection ends in one of two ways, either when a request uses the block or when its timeout expires unused (120s in our runs). In either case, it joins the CACHED queue.

Protection is considered a priority, but the blocks are not pinned. If a demand read or a store finds no unprotected block to evict, it evicts a protected one. A wrong or excessive prefetch therefore degrades L2 toward basic LRU instead of blocking demand traffic. Moreover, a small amount of reserve space (256 blocks or 1.5 GiB in our runs) is kept free on the L2 for blocks in flight between tiers.

\subsubsection{Refill: Watching Blocks Still in L1}
When a prefetch is issued, part of the agent's context may still be in L1. In our experiments, we see that up to 30\% of prompt tokens may be resident in L1 when the corresponding prefetch is issued. Those blocks are useful only while L1 keeps them, and L1 may evict them to admit other requests before this agent uses them. TripleKV introduces a watch list to solve this problem (Fig \ref{fig:queues}). Each block of the prefix that is in L1 but not in L2 is added to a watch list. When a watched block is evicted and no other GPU block holds the same content, TripleKV reads it from L3 into a protected L2 slot. A watch ends either when (1) the request uses the block in L1, (2) the L1 block is evicted and refill is issued, or (3) the TTL expires. This behaves like a victim cache mechanism \cite{jouppi1990victim}.

% could copy them to L2 when the prefetch is issued, copy them out of HBM when L1 evicts them, or read them again from L3 if L1 evicts them.

% \subsection{SSD Queues: Demand Misses First}

% \begin{figure}[t]
%     \centering
%     \includegraphics[width=\columnwidth]{figs/miss_queue.png}
%     \caption{Prefetch reads are promoted to the higher priority miss queue when their request arrives.}
%     \label{fig:intuition}
% \end{figure}

\subsection{Implementation and Overheads}

TripleKV extends vLLM's offloading connector, which already puts a CPU tier and may be configured to put a file-system tier behind the GPU as write-through storage. We add protection classes to the CPU tier, the prefetch engine and watch list to the tier manager, the three-queue thread pool to the file-system tier, and a \textsc{/kv\_cache/prefetch} endpoint to the server. This is about 3000 lines of code.
% This is about 2,100 lines in vLLM's KV-offload package, 480 in the connector and 290 in the endpoint, plus small hooks in the scheduler and the block pool. 
The predictor is implemented in the agentic harness. It is under 300 lines of Python code, and issues one prefetch call per tool call. Our runs use 64-token offload blocks (6 MiB) and 64 GiB of L2 per GPU (about 10,900 blocks) with a 120s timeout.

TripleKV's costs fall outside the critical path. It reserves no GPU memory and adds no GPU work, since blocks reach L1 through vLLM's existing loads from L2. A prefetch call is asynchronous and its reads have lower priority than block reads on the critical path of requests. %It returns once vLLM has claimed the blocks at the next engine step, after 60–380 ms at the median and under 1.3 s at p99, and the agent never waits for it. 
The bookkeeping is a few table entries per protected or watched block that adds up to a few thousand entries at most in total. %The real cost is SSD bandwidth, because prefetches and refills may read blocks that go unused. §7 measures it as prefetch accuracy and coverage.

\section{Evaluation}
\label{sec:eval}
We compare TripleKV with the systems mentioned in Section \ref{sec:baselines}, namely, vLLM (with L3), TokenCake, Continuum (with LMCache) and Oracle on the four workloads described in Section \ref{sec:benchmarks}, at 12, 16 and 20 concurrent agent threads per GPU. A thread picks up a task from the benchmark, runs its agentic loop (with inference and tools) until completion. It picks up a new task from the benchmark once its previous task finishes.

\subsection{Evaluation Methodology}
\label{sec:methodology}

\subsubsection{Platform}

Every experiment run uses one GPU for one serving engine, with the three KV tiers. The SSD tier is large enough and never evicts. The setup serves a 30B mixture-of-experts model. Table \ref{tab:platform} lists the hardware and software configurations. 
\begin{table}[htbp!]
\centering
\small
\caption{Platform and engine configuration.}
\label{tab:platform}
\begin{tabularx}{\columnwidth}{@{}lX@{}}
\toprule
Component & Configuration \\
\midrule
GPU & NVIDIA H100 NVL, 94\,GB HBM\\
Model & Qwen3-30B-A3B~\cite{qwen30b} (60\,GiB in BF16)\\
L1: HBM & 32\,GiB after the model weights\\
L2: DRAM tier & 64\,GiB\\
L3: SSD tier & 1\,TiB hardware RAID of SSDs read with \texttt{O\_DIRECT}\\
Host & Intel Xeon Platinum 8592+ (64 cores), Ubuntu 24.04 \\
Software & vLLM v0.24 (June 2026), PyTorch 2.11, CUDA 13.0\\
\bottomrule
\end{tabularx}
\end{table}

\added{In our evaluations, we assign 64 GiB of DRAM for the KV cache.
We think this size is reasonable for two reasons. First, we want to
emulate a highly utilized system, with many agents, their tools, and processes.
In our experiments, a 50 agent session generates 200–500 GiB of context, which is significantly more than the available host memory. Moreover, the tools and processes also use a significant amount of host memory. 
% These tools and processes also use a significant amount of host memory. Moreover, in our experiments, a 50 agent session generates 200–500 GiB of context, which is significantly more than the available host memory.
Second, we use sparse models, which use less memory per token of context. A comparable-sized dense model would require up to three times the KV Cache space to store the same context, which would make the system much more loaded. We limit host memory to 64 GiB to keep the observations representative for a variety of models and higher load.

% Qwen3-32B~\cite{qwen3-32b}, generates $2.7\times$ the KV per token and needs $2.7\times$ the DRAM. 

% First, a finished session's KV stays useful for the user's next turn or a follow-up task and stays on the DRAM as well.  Second, the working set scales with the model. A comparable dense model, Qwen3-32B~\cite{qwen3-32b}, generates $2.7\times$ the KV per token and needs $2.7\times$ the DRAM. We limit host memory to 64 GiB to keep the observations representative for a variety of models and higher load. %Moreover, the ratio of the working set to the amount of DRAM available is what decides how the system behaves. %At 12–20 agents, 64 GiB holds $0.5-1.4\times$ W, the regime in which that ratio reaches and exceeds one.
}
% \xa{This is not the place for this -- you should move this to methodology. The first part I do not get. The second part I get it a bit, you could write something like: We limited host memory to 64GB to keep the observations representative for larger models and higher load? Keep it short. If you get the chance ask also Josep.}

% In our evaluations, we give the system 64 GiB of DRAM for KV cache, for two reasons. First, a finished session's KV stays useful for the user's next turn or a follow-up task and stays on the DRAM as well. In our experiments, a 50 agent session leaves 200–500 GiB of context, which is significantly more than the available host memory. Second, the working set scales with the model. A comparable dense model, Qwen3-32B~\cite{qwen3-32b}, generates $2.7\times$ the KV per token and needs $2.7\times$ the DRAM. We limit host memory to 64 GiB to keep the observations representative for a variety of models and higher load.
\subsubsection{Benchmarks}

\label{sec:benchmarks}

\begin{figure}[htbp!]
    \centering
    \includegraphics[width=0.9\columnwidth]{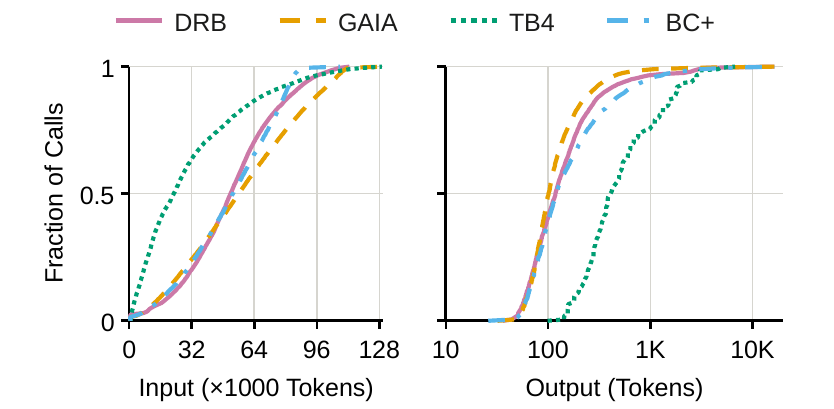}
    \caption{Input and Output Length Distribution}
    \label{fig:benchmarks}
\end{figure}

We use four agent benchmarks as shown in Table \ref{tab:workloads}. Their agents use vastly different tools, so their tool latencies (Fig \ref{fig:tool_latency}) differ. Their input and output lengths (Fig \ref{fig:benchmarks}) differ as well. The workload is input-heavy, as is typical of agentic tasks.

As LLMs can be irregular and multiple runs of the same task can take different trajectories and use different tools, we generate traces for each benchmark. Each trace comes from a live run of the benchmark on vLLM and holds 50 random tasks. We then replay these recorded traces to evaluate our system and baselines fairly to ensure a consistent input size, output size, and tool latency.

\begin{table}[htbp!]
\centering
\small
\caption{Workloads.}
\label{tab:workloads}
\begin{tabularx}{\columnwidth}{@{}l>{\raggedright\arraybackslash}X>{\raggedright\arraybackslash}X>{\raggedright\arraybackslash}X@{}}
\toprule
Benchmark & Agent harness & Tools & Tasks \\
\midrule
DRB~\cite{du2025deepresearchbench} & Tongyi ReAct & Web search, page visits, page summaries & Open-ended research questions \\
GAIA~\cite{mialon2024gaia} & Tongyi ReAct & Web search, page visits, Python sandbox & Everyday, multi-step web skills \\
TB4~\cite{merrill2026terminalbench} & terminus-2 in Harbor & Shell commands & Software tasks in a terminal \\
BC+~\cite{chen2025browsecompplus} & Tongyi ReAct & Local search, document reads & Hard multi-hop questions \\
\bottomrule
\end{tabularx}
\end{table}

% \hf{what is going on with this sentence?}: tool latency, context length and GPU load (Table 5). Each trace comes from a live run with vLLM's tiered LRU. DRB and GAIA have one trace per agent count, recorded at that count. Terminal-Bench 4.0 and BrowseComp-Plus have one trace each, recorded at 20 agents and replayed at every count. We drop SWE-bench: at 16 agents or fewer its contexts fit in memory, and it has no runs on the unified tier.

\subsubsection{Baselines}

\label{sec:baselines}

We evaluate TripleKV against recent state of the art baselines:

\begin{enumerate}[leftmargin=*]
    \item \textit{vLLM \cite{vllm} with L3:} vLLM has become the de facto standard for LLM inference serving. We configure it to use an inclusive, write-through L3 disk tier for KV Cache offloading in our fork.
    \item \textit{TokenCake \cite{tokencake}:} TokenCake on vLLM with its agent-aware scheduling and CPU-saving techniques from its public artifact.
    \item \textit{Continuum \cite{continuum} + LMCache~\cite{lmcache}:} Continuum on vLLM with LMCache for KV Cache management on DRAM from its public artifact.
    \item \textit{Oracle:} TripleKV's mechanism with oracle tool latencies from the trace instead of predictions.
\end{enumerate}

We keep the L2 KV Cache tier fixed at 64 GiB of host DRAM for all the baselines.

\subsection{Latency and Throughput}

\begin{figure}[htbp!]
    \centering
    \includegraphics[width=0.8\columnwidth]{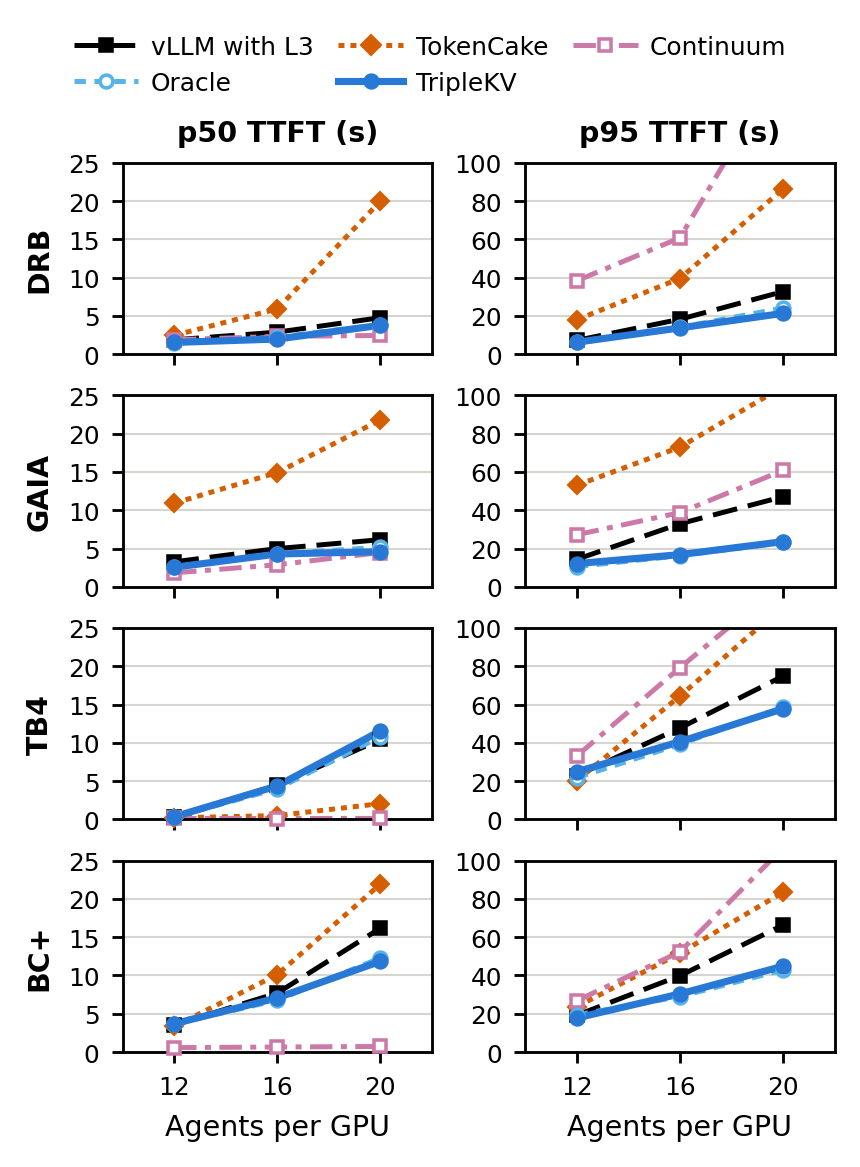}
    \caption{Median and tail TTFT.}
    \label{fig:ttft}
\end{figure}

\paragraph{Time to First Token (TTFT)}
TripleKV has the lowest tail (P95) TTFT on all benchmarks at every degree of concurrency (Figure~\ref{fig:ttft}). Against vLLM with L3 it cuts p95 TTFT by up to 35\%, 49\%, 23\% and 33\% on DRB, GAIA, TB4 and BC+ respectively.
The gains are largest where vLLM reads the most from L3. The median latency on the other hand, remains either comparable or up to 32\% lower across the four benchmarks. This is because the well-performing requests tend to find their KV Cache data on L1 or L2 more often, in the case of vLLM.

TripleKV's latency always remains within 5\% of the Oracle, where we use the actual tool latencies from the trace instead of predicting them. This shows TripleKV’s predictors are accurate enough and that the design tolerates the prediction errors in Fig~\ref{fig:tool_latency}. At a few points TripleKV is marginally below the Oracle, which we attribute to SSD read queuing and run-to-run noise.

TokenCake performs worse than TripleKV on every benchmark, both at median and tail, except for TB4 at median.
This is because TB4 has smaller input sizes for most requests (Fig \ref{fig:benchmarks}) and the median request doesn't need to use L3 at all. Thus, TripleKV's prefetching isn't able to help at the median. However, TripleKV beats TokenCake by up to 77\% consistently at the tail. This is because TokenCake sacrifices the tail performance and queuing to optimize for shorter requests that need less KV cache.

Continuum, on the other hand, consistently proved to be comparable to or better than TripleKV at the median, but sacrifices the tail completely. This is because it pins the KV cache in the GPU memory for running agents. Thus, some agents don't wait at all (thereby improving the median) whereas other wait for very long durations. Therefore, TripleKV beats it by up to 86\%.%\hf{looking at fig14, it doesn't look that Continuum sacrifices the tail at 20 agents, I think you have to comment on that or soften, what would explain that the tail latency goes down for 20? is there more sharing out of L2 in the prefix ?}

\begin{figure}[htbp!]
    \centering
    \includegraphics[width=0.8\columnwidth]{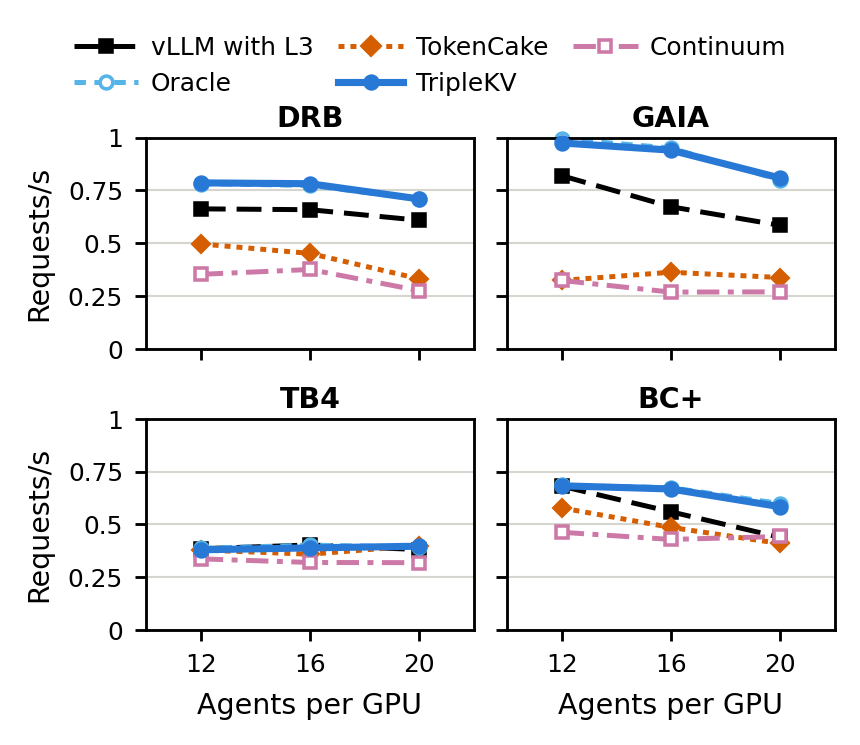}
    \caption{Steady State Request Throughput.}
    \label{fig:req_throughput}
\end{figure}

\paragraph{Throughput.}
TripleKV serves at least as many requests per second as every baseline at every point, as shown in Fig~\ref{fig:req_throughput}. In fact, against vLLM with L3, it serves up to 18\% more on DRB, up to 40\% more on GAIA, up to 31\% more on BC+, and the same on TB4. TokenCake observes up to 2.9$\times$ smaller request throughput against TripleKV consistently on all four benchmarks. This is because it induces long delays in the system due to prefetch requests contending with inference requests for empty KV Cache blocks on the GPU memory. Continuum also observes up to 3.5$\times$ lower request throughput than TripleKV on every benchmark as it prioritizes certain agents by pinning their KV Cache in the GPU memory while neglecting others and forcing them to wait.

\begin{figure}[htbp!]
    \centering
    \includegraphics[width=0.8\columnwidth]{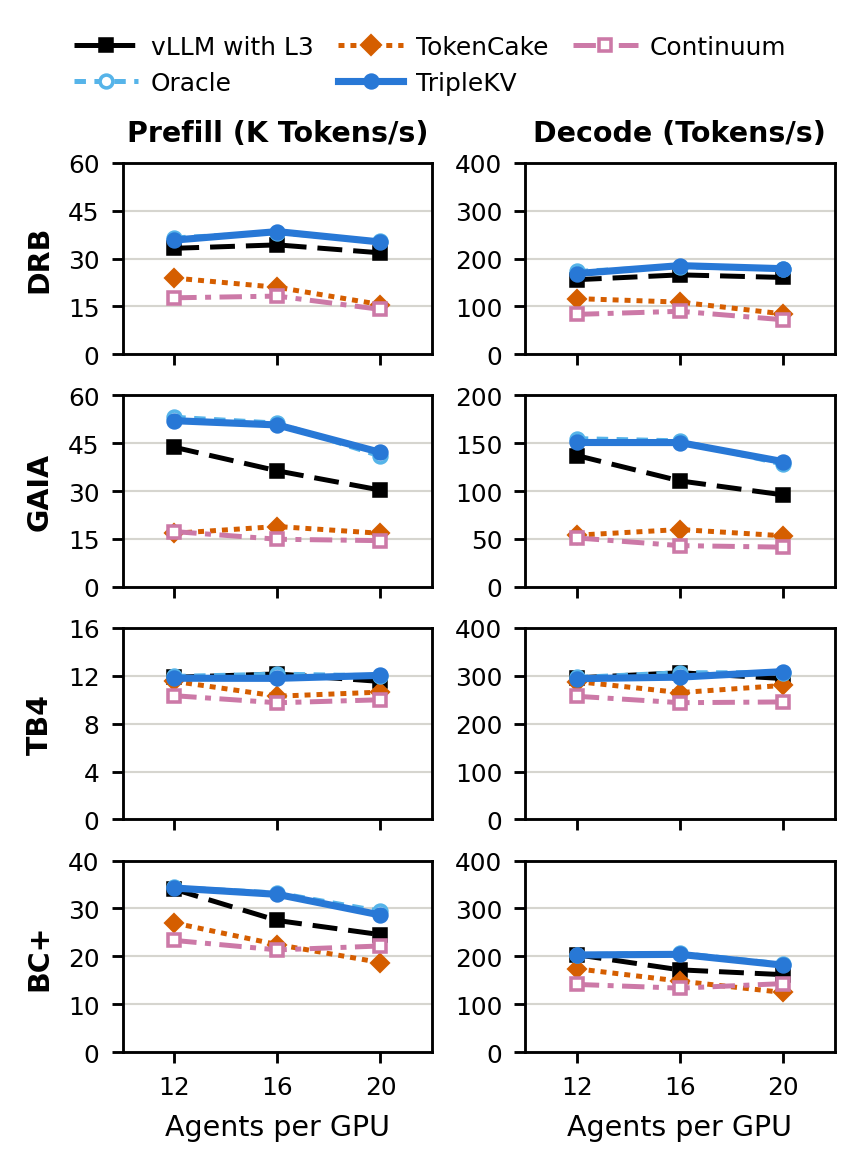}
    \caption{Prefill and Decode Throughput.}
    \label{fig:pd_throughput}
\end{figure}

Fig~\ref{fig:pd_throughput} shows the prefill and decode throughput of the systems. This is measured in tokens per second instead of requests per second. 
Prefill counts every prompt token the engine processed, whether its KV cache came from a tier or was computed. Decode counts every output token. 

The trends remain similar. Against vLLM with L3, TripleKV prefills and decodes up to 40\% more tokens per second across the four benchmarks. Against TokenCake and Continuum, TripleKV prefills and decodes up to 3.5$\times$ as many tokens per second across our benchmarks. We also see that the throughput remains relatively stable as the degree of concurrency increases.
Moreover, TripleKV performs within 5\% of Oracle in throughput as well.%\hf{there is no mentioning of the Oracle before wrt to evaluation.}

\paragraph{Request Delay.}
\begin{figure}[t]
    \centering
    \includegraphics[width=0.9\columnwidth]{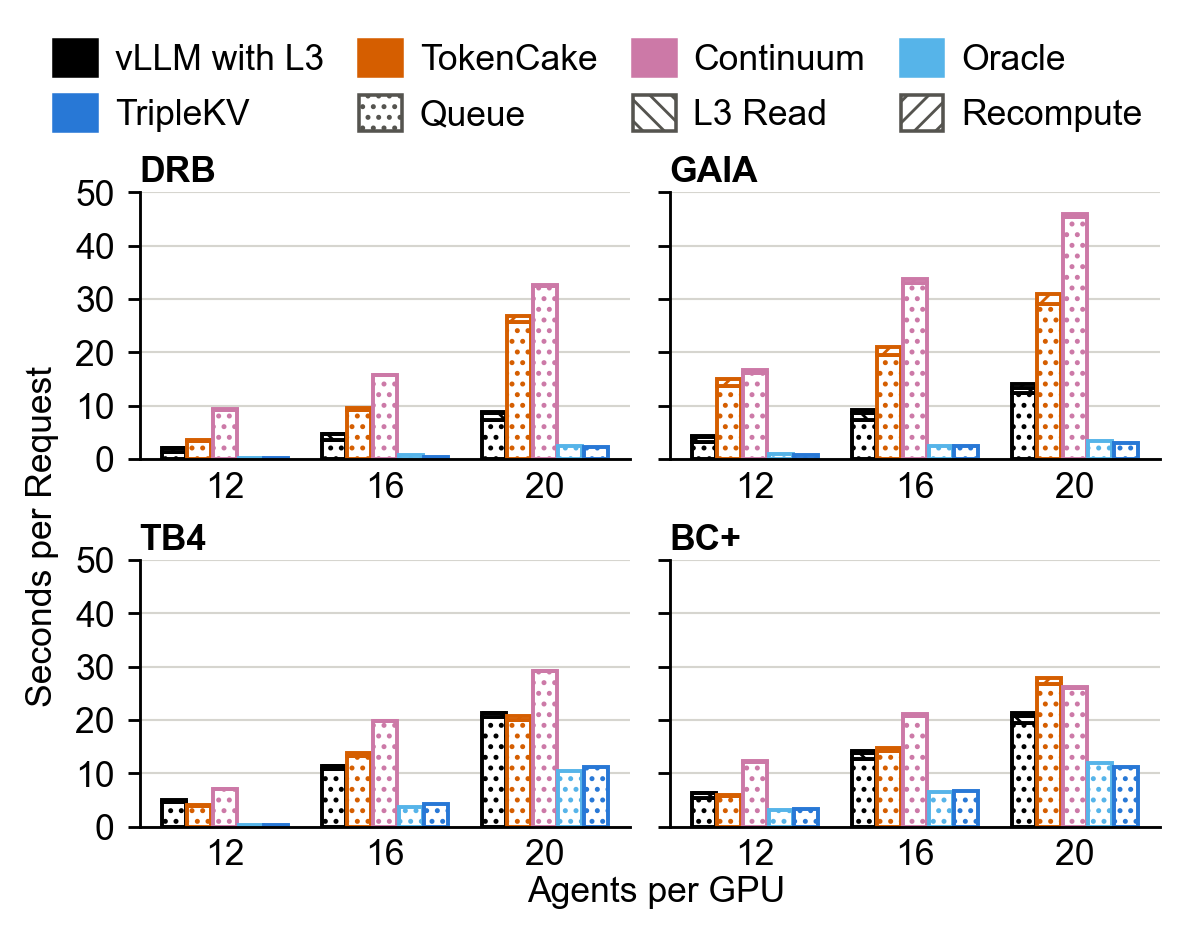}
    \caption{Average Request Delay.}
    \label{fig:delay}
    \vspace{1mm}
\end{figure}

Fig~\ref{fig:delay} splits the mean time a request waits before its prefill starts into three parts: queueing for the GPU, reading its KV Cache from L3, and recomputing context it lost due to evictions. At 20 agents, TripleKV's requests wait up to 78\% less than what they wait on vLLM (with L3). The latter spends up to 1.6\,s per request on average, just reading from L3 whereas TripleKV spends at most 0.05\,s up to 16 agents and at most 0.45\,s at 20, due to its prefetching.

All systems suffer from queuing. However, by Kingman's formula~\cite{kingman1961}, the mean queueing delay grows with the mean request service time, so the time a request holds GPU memory while loading its KV lengthens the queue for every request behind it. Therefore, as TripleKV holds the GPU for the shortest time, it has the least amount of queuing. It is also within 0.5\,s of Oracle.

% The queue for the GPU is the larger part, over 90\% of TripleKV's wait everywhere, and it shrinks more than the read itself: on GAIA at 20 agents, TripleKV's queue is 7.7\,s against vLLM with L3's 12.3\,s. A request that loads its KV at admission holds GPU memory while it waits, and every request queued behind it waits too. The two baselines without an L3 tier show the same effect more strongly. 
TokenCake recomputes for at most 1.9\,s per request, but the recompute occupies the GPU. Therefore, its queue blows up and is up to 11$\times$ TripleKV's at 20 concurrent  agents. Continuum spends the least time on KV loading as it pins it in the GPU. However, its requests queue the longest, up to 15$\times$ TripleKV's in some cases. The agents whose KV it pins in the GPU memory skip the reload, and the rest of the agents wait behind them, thereby increasing the queuing delay. 

TripleKV sees a relatively smaller improvement against the baselines on BC+. This is because the benchmark has long input sizes (Fig \ref{fig:benchmarks}) and short tool calls (Fig \ref{fig:tool_latency}), thereby keeping many large contexts in flight at once and increasing queuing for all systems.

% \subsection{Prefetch Metrics}

\subsection{Cache Metrics}

We evaluate TripleKV's prefetching with standard cache metrics like coverage, precision and timeliness.
% Every metric remains close to ideal. At the two points where demand reaches capacity, GAIA~20 and BC+~20, coverage drops, while precision and timeliness stay high.

% We then check its DRAM capacity. Below DRAM capacity, every metric is close to ideal. At the two points where demand reaches capacity, GAIA~20 and BC+~20, coverage drops, while precision and timeliness stay high.

\paragraph{Cache Coverage.}
\begin{figure}[htbp!]
    \centering
    \includegraphics[width=0.8\columnwidth]{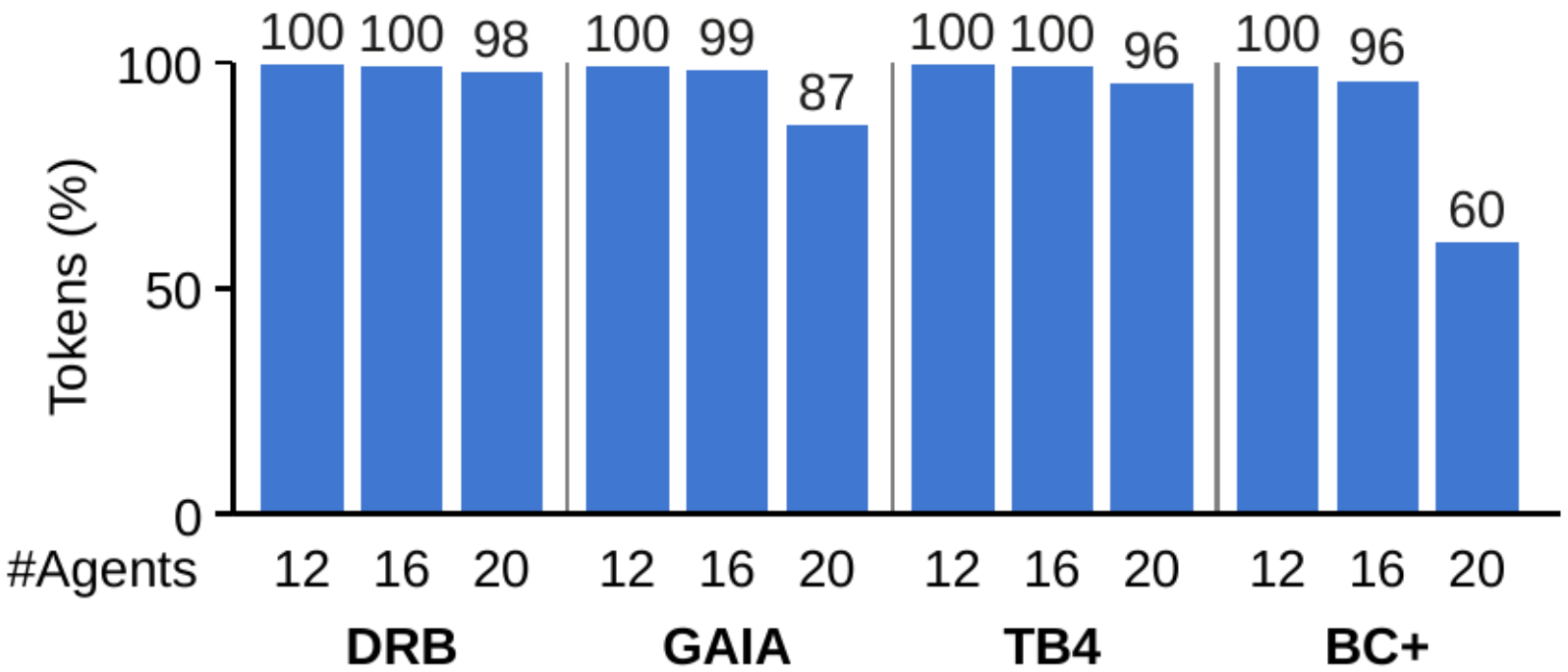}
    \caption{Cache Coverage.}
    \label{fig:coverage}
\end{figure}
We define cache coverage as the share of the KV cache data a request loads from below L1 that it finds in DRAM (L2), thereby avoiding reads from SSD (L3) at admission. Fig \ref{fig:coverage} shows that TripleKV's coverage is 96--100\% at every point except GAIA~20 (87\%) and BC+~20 (60\%). %vLLM with L3 finds only 5--39\% of that KV in DRAM, so 28--76\% of its requests wait on an SSD read. 
% Under TripleKV that share is at most 3.4\% outside GAIA~20 (9.3\%) and BC+~16--20 (5.1\% and 24\%).
This is because, at these two points, the contexts TripleKV protects outgrow the available DRAM capacity. Prefetched blocks are then evicted before their request arrives, 13\% of them on GAIA~20 and 38\% on BC+~20, and the request reads them again from SSD at admission. Below capacity, almost no prefetched block is lost, and coverage stays near 100\%. For comparison, vLLM with L3 finds only 5--39\% of this KV in DRAM.

% : protected demand reaches 63\,GiB and 83\,GiB, against the 62.5\,GiB that prefetches may protect (Fig~\ref{fig:capacity}). Prefetched blocks are then evicted before their request arrives, 13\% of them on GAIA~20 and 38\% on BC+~20, and the request reads them again from SSD at admission. Below capacity, almost no prefetched block is lost and coverage stays near 100\%. For comparison, vLLM with L3 finds only 5--39\% of this KV in DRAM.

\paragraph{Cache Precision.}

\begin{figure}[htbp!]
    \centering
    \includegraphics[width=0.8\columnwidth]{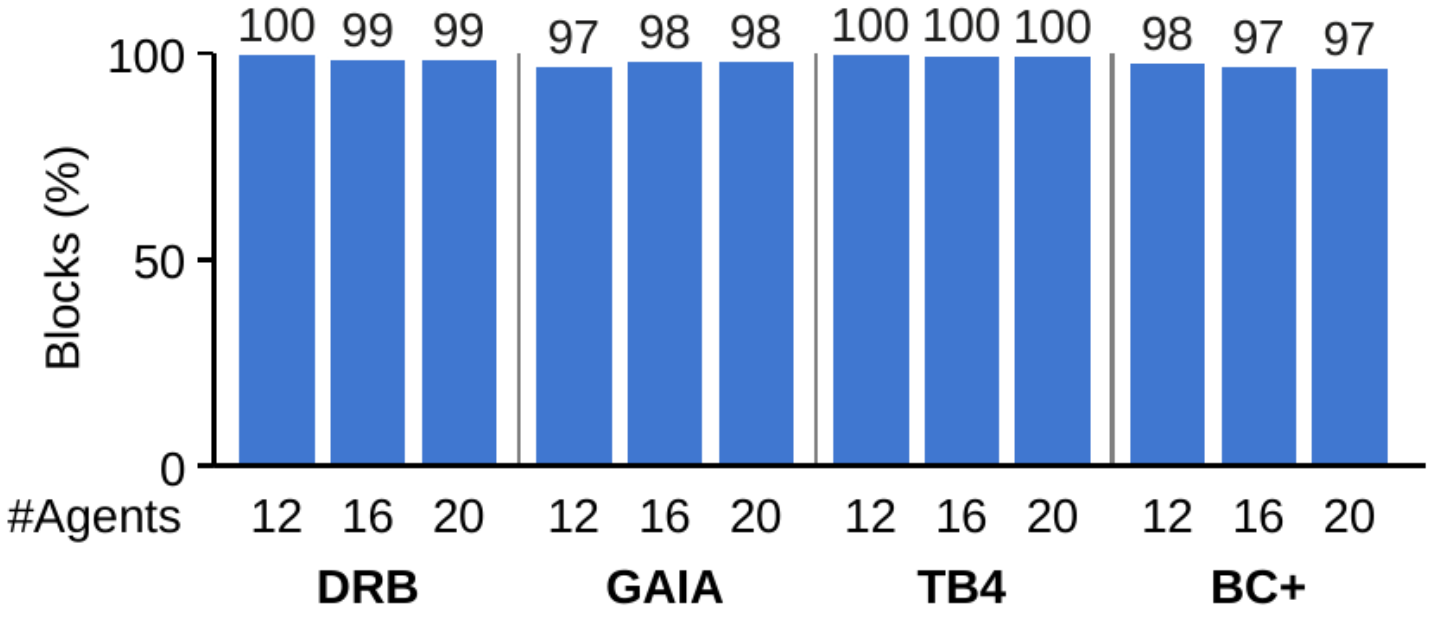}
    \caption{Cache Precision.}
    \label{fig:precision}
\end{figure}
Precision is the share of the blocks a prefetch moved from SSD to DRAM that its request needed. TripleKV's precision is 97--100\% at every point, including the two points where coverage dropped. This is shown in Fig~\ref{fig:precision}.

However, we observe that there can be two kinds of prefetched blocks that remain unneeded. First, the end of the previous prompt. The very last block of a prompt may not be full, and the next request may append to it, thus modifying its hash. Therefore, it will identify as a new block and not match with the old prefetched block. Second, context the agent summarized. Agents can occasionally summarize their context as it grows past a certain threshold to save tokens. After summarization, the entire context must be prefilled again, and the old context's KV Cache is no longer useful. Despite this, precision remains close to 100\%. %As the A block the request needed but that was evicted before it arrived counts as precise here; its cost shows up in coverage and capacity instead.

\begin{figure}[htbp!]
    \centering
    \includegraphics[width=0.8\columnwidth]{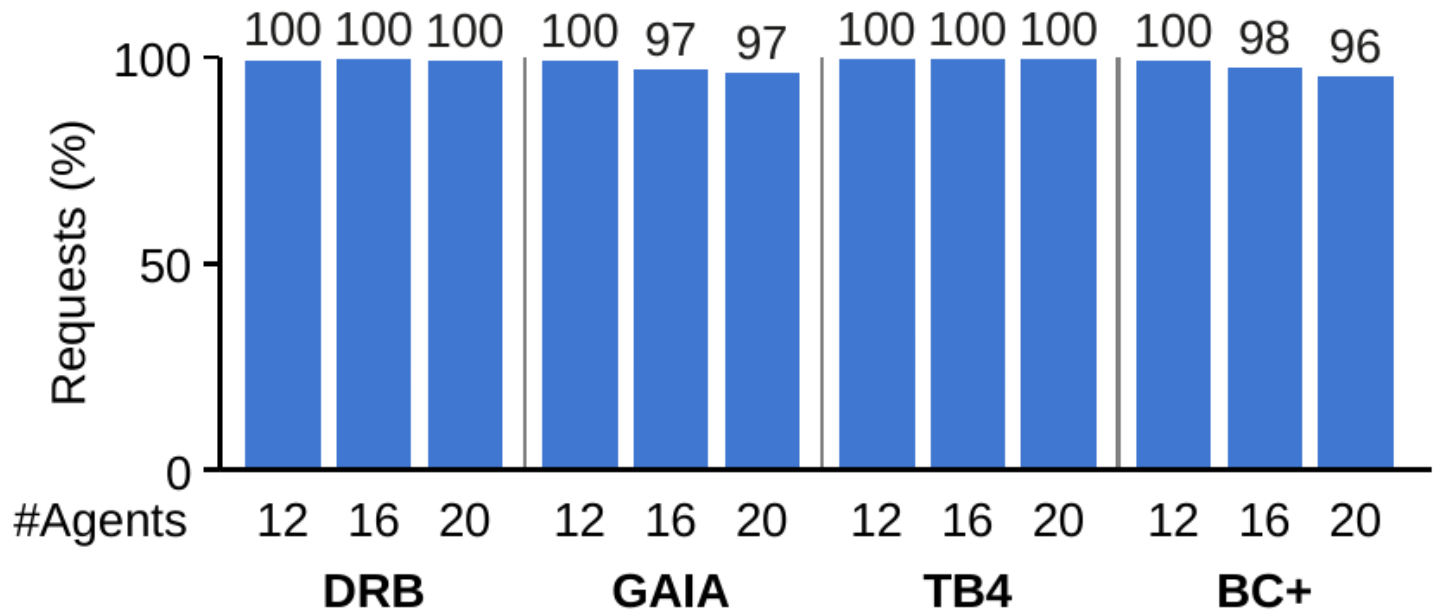}
    \caption{Prefetch Timeliness (requests whose prefetch finished before they were admitted).}
    \label{fig:timeliness}
\end{figure}

\paragraph{Timeliness.}
A prefetch is timely if it finishes before its request is admitted to be scheduled on the GPU. This is when the scheduler reads the KV Cache. TripleKV's prefetches are timely for 96--100\% of requests at every point as shown in Fig~\ref{fig:timeliness}. The predicted tool end is late enough to keep protection short, and early enough to hide the L3 read from the request's critical path. We also observe TripleKV's timeliness remains within 1\% of that of the Oracle runs, which know each tool's end time exactly (not shown).

\paragraph{Cache Capacity.}

\begin{figure}[htbp!]
    \centering
    \includegraphics[width=0.8\columnwidth]{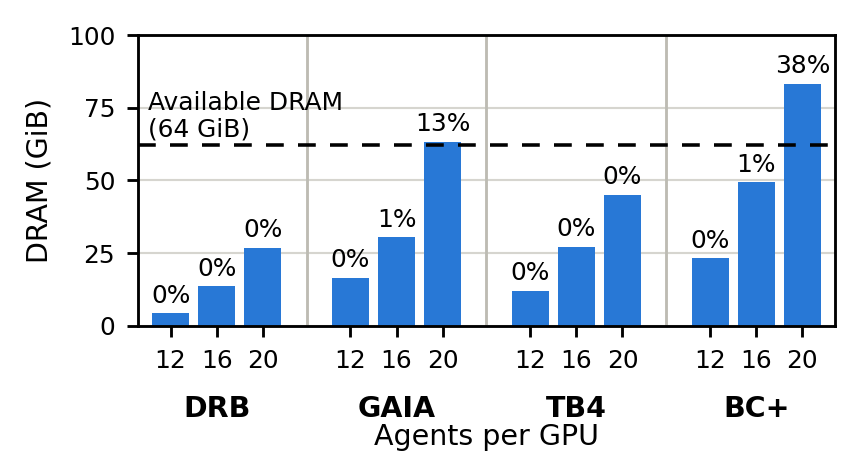}
    \vspace{-1mm}
    \caption{Prefetched memory demand. Above each bar, the share of prefetched blocks evicted before use.}
    \label{fig:capacity}
    %\vspace{1mm}
\end{figure}
Protected demand is the DRAM needed to hold every prefetched context from its prefetch until it can be used. TripleKV's demand grows with the number of concurrent agents running in the system, from 4--23\,GiB at 12 agents to 27--83\,GiB at 20 as shown in Fig~\ref{fig:capacity}. It touches the 64\,GiB of DRAM that is available in the system at GAIA~20 (64\,GiB) and BC+~20 (83\,GiB). As mentioned earlier, about 2 GiB space is kept reserved for in-flight blocks between L1 and L3 and cannot be used to store prefetched blocks. Thus, in these two experiments, we observe some prefetched blocks getting evicted before their request is admitted. This is about 13\% and 38\% of the blocks, respectively, as shown on top of the bars in Fig \ref{fig:capacity}. These evictions cause the drop in cache coverage illustrated earlier.

\vspace{-2mm}

\subsection{Ablation Study}
\begin{figure}[t]
    \centering
    \includegraphics[width=0.8\columnwidth]{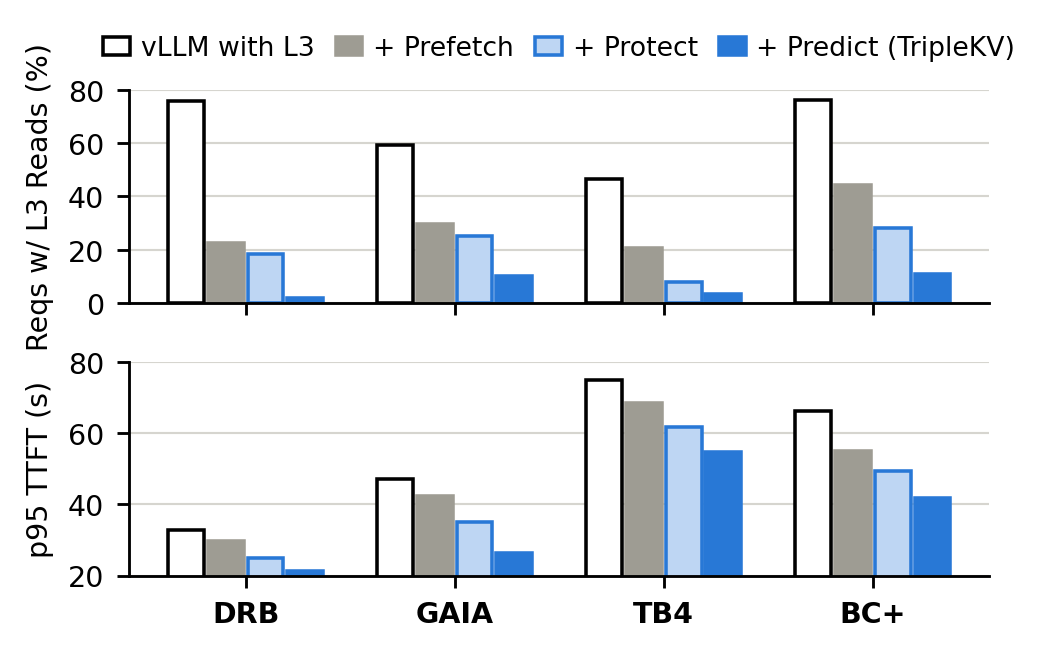}
    \vspace{-1mm}
    \caption{Ablation Study with 20 Concurrent Agents}
    \label{fig:ablation}
    \vspace{1mm}
\end{figure}

Fig \ref{fig:ablation} shows the ablation study when we incrementally add the three components of TripleKV. We start from a basic vLLM (with L3) design and add prefetching to this. Here, the prefetches are issued naively as soon as the tool starts. Next, we add the protect component to it that makes sure that the prefetches stay on L2 until they are used. Finally, we add the Predict component that issues the prefetches exactly when they are needed.%\hf{can you state that these are incremental improvements. Looks to me that having more capacity is the most important thing (vllm+L3) as it avoid recomputation, based on that figure prediction doesn't add much.}

The top figure shows the fraction of request that need to read from L3 in their critical path. We observe that with every added component, the share drops. Naive prefetching still leaves 21–45\% of requests reading from L3 on their critical path as the prefetched blocks get evicted before the request arrives. Adding the protect component helps, but protecting too many blocks too early reduces the available DRAM space for other requests. Finally, adding the predict component prefetches when blocks are needed and protects them in DRAM only for the required time. The tail latency (bottom) also follows similar trend and improves with every added component.
\section{Related Work}
\label{sec:related}
% Existing keys used below map to the current reference list:
% infercept=[1] symphony=[2] nalar=[3] belady=[4] tokencake=[5] cao95=[6]
% cachedattention=[7] hawkeye=[9] continuum=[13] lmcache=[14] askthetool=[15]
% kvbm=[16] kvflow=[17] tip=[18] mooncake=[19] mockingjay=[21] mori=[23]
% strata=[24] pbkv=[27]. Rename to your own keys.

\textit{Agent-aware KV management.}
Systems handle an idle agent's KV in three ways. \emph{(1) Prefetching:} TokenCake~\cite{tokencake} reloads stalled agents from DRAM into HBM before predicted completion. Its predictions are based on a moving average of the tool latency and developer hints. KVFlow~\cite{kvflow}, PBKV~\cite{pbkv} and CacheScout~\cite{zhang2026cachescout} prefetch the context of the agents that is predicted to run next, without considering when it will run.
\emph{(2) Retention:} Continuum~\cite{continuum} pins KV in HBM for a time-to-live determined by reload and queueing costs. InferCept~\cite{infercept} keeps, swaps or recomputes paused requests' KV to minimize idle memory waste, and MORI~\cite{mori} keeps only busy agents in HBM based on their recent idleness. \emph{(3) Application cues:} SYMPHONY~\cite{symphony} and Nalar~\cite{nalar} act on application hints and Ask-the-Tool~\cite{asktool} on tool-reported progress to perform KV cache movement between DRAM and GPU. These require reprogramming the application or its tools, which is a significant overhead. TripleKV adds a predictor only to the agent harness and works with existing tools or the application. It prefetches KV cache blocks into the host DRAM and protects prefetched blocks to prevent premature eviction.
% \paragraph{Agent-aware KV management.}
% Agent-aware systems differ in the tier they manage and the signal that is used to perform each data movement.
% Continuum~\cite{continuum} pins KV in HBM for a TTL determined by reload cost and queueing delay.
% TokenCake~\cite{tokencake} offloads stalled agents to host DRAM, reloads them into HBM before predicted completion, and reserves HBM for critical agents. 
% InferCept~\cite{infercept} keeps, swaps or recomputes each paused request's KV cache data.
% MORI~\cite{mori} offloads by idleness phase, Ask-the-Tool~\cite{asktool} reads live tool progress, and SYMPHONY~\cite{symphony} and Nalar~\cite{nalar} act on application hints.
% KVFlow~\cite{kvflow} and PBKV~\cite{pbkv} prefetch predicted agents' prefixes from host memory into the GPU; CacheScout~\cite{zhang2026cachescout} learns such predictions online.
% Most target the HBM--DRAM boundary.
% TripleKV applies the tool-completion signal one level lower: it stages contexts from SSD into DRAM, never reserves HBM, and protects staged blocks until admission.

\textit{Hierarchical KV caching.}
CachedAttention~\cite{cachedattention} and Strata~\cite{strata} fetch KV from disk into DRAM after the requests have been queued, and CachedAttention exempts its look-ahead window from eviction.
LMCache~\cite{lmcache}, Mooncake~\cite{mooncake}, SGLang~\cite{sglang}, HiCache~\cite{xie2025hicache} and Dynamo KVBM~\cite{dynamo} extend the cache to storage. HiCache fetches into host memory after the request is known once its prefix match passes a threshold.
Pensieve~\cite{yu2025pensieve} swaps multi-turn chat state between GPU and CPU ahead of time. DualPath~\cite{wu2026dualpath} finds agentic inference storage-bound and proposes reading the KV cache through two network paths instead of one.
These system start reading once the request is known.
TripleKV reads the offloaded KV cache before the agent resumes and hides the read latency during the tool execution phase.% at the predicted tool end and exploits that the upcoming request's prefix is already known by virtue of it being part of an agentic request.

% \paragraph{Informed prefetching and integrated caching.}
% TIP~\cite{patterson} prefetches from disclosed future accesses and bounds how far ahead to issue them.
% Cao et al.~\cite{cao} give rules for when to prefetch and what to evict, which Kimbrel and Karlin~\cite{kimbrel2000parallel} and Albers et al.~\cite{albers2000stall} extend to parallel disks and optimal stall time.
% TripleKV's issue rule (Eq.~\ref{eq:issue}) is a deadline form of TIP's horizon, with a tool predictor in place of disclosure; its timing error is why staged blocks need protection.

\textit{Informed prefetching and integrated caching.}
TIP~\cite{patterson} prefetches from disclosed future accesses and bounds how far ahead to issue them.
The Aggressive and Conservative policies~\cite{cao} set rules for when to prefetch and what to evict. Reverse Aggressive~\cite{kimbrel2000parallel} extends them to parallel disks, and LP-based schedules~\cite{albers2000stall} minimize stall time optimally.
TripleKV's prefetching is a deadline form of TIP's horizon, with a tool predictor in place of disclosure. %Protection is needed to account for system variability and timing errors.

\textit{Prefetch-aware replacement and exclusive hierarchies.}
TripleKV's Protect follows prefetch-aware replacement in CPU caches~\cite{wu2011pacman, zhou2001mq} and GPU unified memory~\cite{ganguly2019uvm}; like MIN-based policies~\cite{Belady,hawkeye,mockingjay}. It uses knowledge of future reuse.
The watch list resembles victim caching~\cite{jouppi1990victim} and DEMOTE-style exclusive caching~\cite{wong2002demote}, but it refills evicted blocks from the lower tier instead of capturing them on eviction, to avoid stalling the GPU on each eviction.
% OS memory tiering such as TPP~\cite{maruf2023tpp} promotes pages after they turn hot whereas TripleKV promotes contexts before their predicted next use.

\vspace{-2mm}
\section{Conclusion}

This paper presented {\em TripleKV}, a novel agent-aware KV cache management framework.
TripleKV predicts when each tool will finish, prefetches the agent's KV blocks from SSD into DRAM before the next request, and protects them there until they are used. 
Across four agentic benchmarks,
TripleKV reduces the fraction of requests requiring critical-path SSD reads from 45--76\% to under 10\%. It also reduces P95  TTFT by up to 49\% compared with vLLM extended with DRAM and SSD KV-cache tiers, and achieves  1.5-7$\times$ lower P95 TTFT than recent agent-aware systems.

\begin{comment}

In agentic workloads, an agent's KV cache is pushed out of fast memory while it waits on a tool, so its next call reloads that context from a slower medium or recomputes it. TripleKV predicts when each tool will finish, prefetches the agent's KV from SSD into DRAM before the next request, and protects it there until it is used. On four agentic benchmarks, this cuts requests that wait on SSD by 15$\times$ on average. Tail TTFT latency also drops by up to 7$\times$ against baselines. Extending TripleKV across nodes with shared storage is a future work.

% \hf{I am confused by the Acknowledgment}

\end{comment}

% \begin{acks}
% Acknowledgments and funding go here (allowed in the preprint).
% \end{acks}

\bibliographystyle{ACM-Reference-Format}
\bibliography{refs}

\end{document}